\documentclass[superscriptaddress,showpacs,amssymb,10pt,reprint,aps,prd,longbibliography,nofootinbib,floatfix]{revtex4-1}

\usepackage{graphicx,epsfig,amssymb,times} 
\usepackage{amsmath,amsfonts}
\usepackage{bm}
\usepackage{epstopdf}
\usepackage{hyperref}
\usepackage[caption=false]{subfig}
\usepackage[usenames]{color}   
\usepackage[dvipsnames]{xcolor}

\usepackage[normalem]{ulem}

\definecolor{coolblack}{rgb}{0.0, 0.18, 0.39}
\definecolor{darkred}{rgb}{0.5,0,0}
\definecolor{darkgreen}{rgb}{0,0.5,0}
\definecolor{darkblue}{rgb}{0,0,0.5}
\definecolor{lapislazuli}{rgb}{0.15, 0.38, 0.61}
\definecolor{venetianred}{rgb}{0.78, 0.03, 0.08}
\definecolor{bleudefrance}{rgb}{0.19, 0.55, 0.91}
\definecolor{dogwoodrose}{rgb}{0.84, 0.09, 0.41}
\hypersetup{colorlinks=true, citecolor=darkblue, linkcolor=darkblue, 
urlcolor = darkblue}

\newcommand\numberthis{\addtocounter{equation}{1}\tag{\theequation}}

\begin{document}

\title{\large Charged particles and fields in bumblebee gravity: \\ Periodic orbits and superradiance}
	
\author{Marco A. A. de Paula}
\email{marcodepaula@ufpa.br}
\affiliation{Faculdade de Ciências Naturais, Universidade Federal do Par\'a, Campus Universit\'ario do Tocantins-Cametá, 68400-000, Camet\'a, Par\'a, Brazil.}
\affiliation{Departamento de Física, Universidade Federal da Paraíba, Caixa Postal 5008, 58051--970, João Pessoa, Paraíba,  Brazil.}

\author{Renan B. Magalhães}
    \email{renanbatalha@if.uff.br}
	\affiliation{Instituto de Física, Universidade Federal Fluminense, Niterói, Rio de Janeiro, 24210-346, Brazil.}

\author{V. B. Bezerra}
\email{valdir@fisica.ufpb.br}
\affiliation{Departamento de Física, Universidade Federal da Paraíba, Caixa Postal 5008, 58051--970, João Pessoa, Paraíba,  Brazil.}

\author{A. A. Ara\'{u}jo Filho}
\email{dilto@fisica.ufc.br}
\affiliation{Departamento de Física, Universidade Federal da Paraíba, Caixa Postal 5008, 58051--970, João Pessoa, Paraíba,  Brazil.}
\affiliation{Departamento de Física, Universidade Federal de Campina Grande Caixa Postal 10071, 58429-900 Campina Grande, Paraíba, Brazil.}
\affiliation{Center for Theoretical Physics, Khazar University, 41 Mehseti Street, Baku, AZ-1096, Azerbaijan.}

\begin{abstract}

We investigate the dynamics of charged particles and fields in the background of a charged black hole in bumblebee gravity, aiming to better understand the role played by the Lorentz-violating parameter $l$. We first review the derivation of the charged bumblebee black hole and investigate its main physical and geometrical properties. We then obtain the circular and bound trajectories of charged particles and find that the innermost stable circular orbit increases with $l$. By using a numerical approach, we compute the total absorption cross section for arbitrary values of the scalar wave frequency $\omega$. Since the spacetime is not asymptotically flat, we construct the total absorption cross section by taking into account the modified asymptotic momentum and flux. Analytical approximations are derived in the low- and high-frequency regimes. The low-frequency expression displays a transition velocity that separates two distinct asymptotic branches and generalizes the corresponding Reissner-Nordström result. Our numerical results for the cross section agree with the analytical approximations in their corresponding limits, and the absorption parameter space can be divided into three regions: Unbounded absorption, bounded absorption, and bounded superradiance. In particular, superradiance occurs whenever $\omega < \omega_{c}$, where $\omega_{c}$ is the threshold for superradiance. Moreover, the superradiance amplification decreases as we increase $l$, and charged bumblebee black holes with $l < 0$ lead to greater superradiant scattering than in the Reissner–Nordström case. We also note that although increasing $l$ lowers the potential barrier, it reduces the total absorption cross section because of the geometrical factor $1/(1+l)$ present in the cross section formula.

\end{abstract}

\date{\today}

\maketitle

\section{Introduction}

Despite its theoretical and experimental robustness~\cite{Collaboration2025GWTC40AI,EventHorizonTelescope:2019dse,EventHorizonTelescope:2022wkp}, general relativity (GR) fails to provide a complete picture for the gravity panorama~\cite{wald2010general,Hawking:1970zqf,Hawking:1975vcx}. In particular, the unification of GR with
Quantum Mechanics persists as one of the greatest unsolved problems in Physics. Consequently, efforts have been made to develop alternatives to the standard GR. The models in this landscape of extensions of GR often predict that characteristic low-energy ``relic signatures'' should arise, such as the violation of spacetime symmetries, leading to measurable deviations from predictions of GR~\cite{mattingly2005modern,liberati2009lorentz,berti2015testing}.

Even though deviations from established physics arising from an underlying theory, such as string theory~\cite{KosteleckySamuel1989a,KosteleckySamuel1989b} or noncommutative field theories~\cite{carroll2001noncommutative}, are expected to be suppressed by a large energy scale (such as
the Planck mass), progress can be made through effective field theory descriptions, such as the Standard Model Extension \cite{ColladayKostelecky1998,Kostelecky2004,BaileyKostelecky2006}. In this framework, the vacuum expectation
value of tensor fields is responsible for the violation of the local Lorentz symmetry. 
In the context of gravity, a promising approach to investigate Lorentz violation is the Einstein-bumblebee gravity. This alternative theory of gravity is a vector--tensor realization of spontaneous Lorentz symmetry breaking (therefore also implying the spontaneous violation of
diffeomorphism invariance~\cite{BluhmKostelecky2005}), being characterized by a dynamical vector $B_\mu$ non-minimally coupled to the curvature. The presence of a self-interaction potential drives (spontaneously) the dynamic evolution of the vector field toward a nonzero vacuum expectation value. Therefore, while the field equations remain covariant, a solution of this model may select a vacuum direction and transfer this information to the spacetime metric \cite{BluhmKostelecky2005,BluhmFungKostelecky2008,BluhmEtAl2008,BertolamiParamos2005}.

The first spherically symmetric black hole (BH), in the context of Einstein-bumblebee gravity, used for phenomenological applications possesses a radial spacelike vacuum configuration, namely $\langle B_\mu\rangle =(0,b(r),0,0)$~\cite{CasanaEtAl2018}. Similar bumblebee configurations were also considered in other spherically symmetric scenarios, for instance for including a cosmological constant~\cite{MalufNeves2021,DingEtAl2023}, non-commutativity scenario \cite{AraujoEtAlNoncomm2026} and in the metric-affine extension of the bumblebee model~\cite{AraujoFilhoEtAl2023,AraujoFilho:2024ykw}. The search for non-vacuum solutions of the Einstein-bumblebee model is a challenging task, since the non-minimally coupled terms of the bumblebee vector to the curvature tensor can, in the vacuum state, act as geometrical constraints~\cite{lessa2025self}. As a consequence, the naive search for spherically symmetric solutions of the Einstein-bumblebee gravity sourced by an electromagnetic field results in an unsolvable system of differential equations.

However, recently it has been shown that the non-minimal coupling of Lorentz-violating (LV) fields to the electromagnetic fields, through a term $\gamma B^\alpha B_\alpha F^{\mu\nu}F_{\mu\nu}$, where $\gamma$ is a coupling constant and $F_{\mu\nu}$ is the Faraday tensor, can source spherically symmetric solutions if the coupling constant $\gamma$ has a suitable relation with the coupling constant controlling the non-minimal coupling to the curvature~\cite{Liu:2024axg}. A similar LV coupling to the electromagnetic field was considered for the Kalb-Ramond gravity~\cite{duan2024electrically,Gonzalez:2026gye} and to a (phantom) scalar field in the context of bumblebee and Kalb-Ramond gravity~\cite{magalhaes2026nonminimal}.

These solutions have gained increasing attention, and several observables have been used to examine whether such deformations can be distinguished from their general-relativistic counterparts, using both particles and fields. For instance, regarding the neutral solutions, the study of gravitational lensing, thermodynamics, quasinormal modes, and Hawking emission has been considered~\cite{OvgunJusufiSakalli2018,LiOvgun2020,GomesMalufAlmeida2020,OliveiraDantasAlmeida2021,GogoiGoswami2022,UniyalKanziSakalli2023,LiuFangJingWang2024,AraujoFilho:2025zaj,Shi:2025plr,AraujoFilho2025,AraujoFilho2026}. Concerning the charged bumblebee BHs, the investigation of particle trajectories, optical images, perturbative spectra, greybody factors, and gravitational waveforms has already been considered~\cite{XuEtAlCharged2025,SinghEtAl2025,LiEtAlChargedQNM2025,ShiEtAl2026}. Nevertheless, to the best of our knowledge, the dynamics of charged particles and fields in a charged bumblebee BH spacetime have not yet been investigated in detail.

In particular, the study of the motion of charged particles in charged spacetimes is of interest, since the compact objects that we expect to observe are surrounded by accretion disks, composed of charged particles. Moreover, BHs can harbor a small amount of charge in astrophysical BH systems~\cite{Zajacek:2018vsj,Levin:2018mzg}, which can significantly affect the dynamics of charged particles in the BH surroundings~\cite{Schroven:2017jsp}. In GR, charged particles have been extensively investigated~\cite{grunau2011geodesics,kan2022bound,tursunov2016circular,zhao2018static,schroven2021innermost}. A characterization of charged orbits in charged bumblebee spacetimes is therefore motivated. In particular the study of bound orbits, including both circular and non-circular charged motion.

The wave picture scenario is also interesting. A wave impinging on an absorbing system does not always return with a smaller amplitude. If the absorbed part carries negative conserved energy with respect to the generator of the system, the outgoing flux exceeds the incoming one. Zel'dovich identified this mechanism for a dissipative rotating body \cite{Zeldovich1971,Zeldovich1972}, and its gravitational counterpart was clarified through the Penrose process and the wave analyses of Misner, Starobinsky, and Churilov \cite{PenroseFloyd1971,Misner1972,Starobinsky1973,StarobinskyChurilov1974}. For a Kerr BH, the horizon flux of a bosonic mode changes sign in the range $0<\omega<m\Omega_+$, where $\omega$ is the frequency of the scalar wave, with azimuthal number $m$, and $\Omega_{+}$ is the angular velocity of the BH at the outer horizon. In this context, the wave undergoes scattering with more energy than it originally possessed, which characterizes superradiant scattering.

However, the relevant potential need not originate from rotation. When a scalar field of charge $q$ scatters from a static BH of electric potential $\phi_+$, the frequency measured by the horizon generator is $\omega-q\phi_+$. The transmitted flux is negative for $0<\omega<q\phi_+$, provided that $q\phi_+>0$, and the electrostatic energy of the background is transferred to the reflected wave \cite{Bekenstein:1973mi,DiMenzaNicolas2015,Benone:2015bst}. This effect is especially visible in the total absorption cross section (ACS). For a charged massive scalar field, superradiant modes can drive the total cross section below zero, while the competition between gravitational attraction and electrostatic repulsion determines whether its low--frequency magnitude remains finite or diverges~\cite{Benone:2015bst,Richarte:2021fbi,dePaula:2024xnd,dePaula:2025kif}. These properties make charged superradiance sensitive to modifications of the background geometry, even in spherically symmetric spacetimes due to the Lorentz force. Therefore, LV theories provide an interesting testing ground for exploring alternative theories of gravity through the process of superradiance. For charged bumblebee BHs~\cite{Liu:2024axg}, the radial metric approaches a conical instead of a Minkowski form, so the asymptotic wave number and the relation between flux and scattering amplitudes must be derived for this spacetime. Moreover, the LV parameter modifies the allowed charge of the BH and the potential $\phi_+$ that fixes the superradiant threshold. A consistent treatment must incorporate these effects before the amplification can be compared with the standard charged case.

Driven by these motivations, we investigate the dynamics of charged particles and fields in the background of a charged BH in bumblebee gravity, aiming to better understand the role played by the LV parameter $l$. The remainder of this paper is organized as follows. Sec.~\ref{sec:lqgbh} presents the charged bumblebee BH geometry. Charged particle motion is treated in Sec.~\ref{sec:mcp}, and the scalar radial equation and flux condition are derived in Sec.~\ref{sec:sw}. Sec.~\ref{sec:asw} gives the partial absorption cross sections and their analytical limits. The numerical analysis is reported in Sec.~\ref{sec:mr}, followed by the conclusions in Sec.~\ref{sec:remarks}. Further analyses supplementing the main results are presented in the Apps.~\ref{appx2} and~\ref{appx}. We use the metric signature $(-,+,+,+)$ and set $G=c=\hbar=1$.

\section{Charged BHs in bumblebee gravity}\label{sec:lqgbh}

In this section, starting from the action, we review the derivation of the charged BH obtained in the bumblebee gravity presented in Ref.~\cite{Liu:2024axg}. We also discuss the main properties of this geometry, analyzing, e.g., the metric function.


\subsection{The general remarks}\label{subsec:frame}

The bumblebee model introduces a vector field $B_{\mu}$ whose nonvanishing vacuum expectation value selects a preferred direction in spacetime, leading to the spontaneous breaking of Lorentz symmetry \cite{KosteleckySamuel1989a,KosteleckySamuel1989b,ColladayKostelecky1998,Kostelecky2004,BaileyKostelecky2006,BluhmKostelecky2005,BluhmFungKostelecky2008,BluhmEtAl2008}. In the presence of the electromagnetic sector considered here, the corresponding action can be written as \cite{Liu:2024axg}
\begin{align}
S={}&\int \mathrm{d}^{4}x\sqrt{-g}\left[
\frac{1}{2\kappa}
\left(R-2\Lambda+\xi B^{\mu}B^{\nu}R_{\mu\nu}\right) \nonumber \right. \\
& \left.  
-\frac{1}{4}B_{\mu\nu}B^{\mu\nu}
-V(X)
+\mathcal{L}_{\mathrm{em}}
\right],
\end{align}
with the electromagnetic Lagrangian being given by
\begin{equation}
\label{eq:bumblebee-em}\mathcal{L}_{\mathrm{em}} = \dfrac{1}{2\kappa} \left(1+\gamma B^{\rho}B_{\rho}\right) F^{\mu\nu}F_{\mu\nu},
\end{equation}
where $\kappa=8\pi G$, $\xi$ denotes the nonminimal coupling between the bumblebee field and curvature, and $\gamma$ controls the coupling between the bumblebee and electromagnetic sectors. The corresponding field-strength tensors are
\begin{equation}
B_{\mu\nu}=\partial_{\mu}B_{\nu}-\partial_{\nu}B_{\mu},
\qquad
F_{\mu\nu}=\partial_{\mu}A_{\nu}-\partial_{\nu}A_{\mu}.
\end{equation}
The potential depends on $X=B^{\mu}B_{\mu}\pm b^{2}$ and possesses a minimum at $X=0$. Furthermore, the bumblebee field acquires the vacuum value $\langle B_{\mu}\rangle=b_{\mu}$, with $b^{\mu}b_{\mu}=\mp b^{2}$ \cite{BluhmKostelecky2005,BluhmFungKostelecky2008,BertolamiParamos2005}.

Variation of the action with respect to $g^{\mu\nu}$, $B_{\mu}$, and $A_{\mu}$ gives, respectively,
\begin{align}
& G_{\mu\nu}+\Lambda g_{\mu\nu}
=\kappa\left(T_{\mu\nu}^{B}+T_{\mu\nu}^{\mathrm{em}}\right),
\label{eq:einstein-bumblebee}
\\
&\nabla_{\mu}B^{\mu\nu}
-2\left(
V'B^{\nu}
-\frac{\xi}{2\kappa}B_{\mu}R^{\mu\nu}
-\frac{\gamma}{2\kappa}B^{\nu}
F^{\alpha\beta}F_{\alpha\beta}
\right)
=0,
\label{eq:bumblebee-eom}
\\
&\nabla_{\mu}\left[
\left(1+\gamma B^{\alpha}B_{\alpha}\right)F^{\mu\nu}
\right]
=0.
\label{eq:modified-maxwell}
\end{align}
where $V'=\mathrm{d}V/\mathrm{d}X$, while $T_{\mu\nu}^{B}$ and $T_{\mu\nu}^{\mathrm{em}}$ denote the contributions of the bumblebee and electromagnetic fields to the energy-momentum tensor.

For the charged BH examined below, we set $\Lambda=0$ and adopt the smooth quadratic potential $V(X)=\lambda X^{2}/2$. At its minimum, $B_{\mu}=b_{\mu}$ and $V=V'=0$. Considering the static and spherically symmetric ansatz
\begin{equation}
\mathrm{d}s^{2}=-A(r)\mathrm{d}t^{2}+S(r)\mathrm{d}r^{2}+r^{2}\mathrm{d}\Omega^{2},
\end{equation}
with $b_{\mu}=\left(0,b\sqrt{S(r)},0,0\right)$ and $A_{\mu}=\left(-\phi(r),0,0,0\right)$; in addition, we have $b_{\mu}b^{\mu}=b^{2}$ and $\langle B_{\mu\nu}\rangle =0$ \cite{BertolamiParamos2005,CasanaEtAl2018,Liu:2024axg}. Defining the LV parameter as $l=\xi b^{2}$ and taking $\gamma=\xi/(2+l)$, the independent field equations imply
\begin{equation}
\big[A(r)S(r)\big]'=0,
\qquad
\big[r^{2}\phi'(r)\big]'=0.
\end{equation}
The integration constant in the first relation is chosen as $A(r)S(r)=1+l$, whereas the second relation determines the Coulomb--like potential and its modified charge normalization. These equations lead directly to the charged bumblebee geometry presented in the next subsection.


\subsection{The charged bumblebee black hole}\label{subsec:cbhg}

By solving the field equations, we can show that \cite{Liu:2024axg}
\begin{equation}
\label{LE}\mathrm{d}s^{2} = -f(r)\mathrm{d}t^{2}+\dfrac{1+l}{f(r)}\mathrm{d}r^{2}+r^{2}\mathrm{d}\Omega^{2},
\end{equation}
where $\mathrm{d}\Omega^{2} = \mathrm{d}\theta^{2}+\sin^{2}\theta \,\mathrm{d}\varphi^{2}$ is the line element of a two-dimensional unit sphere, and the metric function reads \cite{Liu:2024axg}
\begin{equation}
f(r) = 1-\dfrac{2M}{r}+\dfrac{(2+l)Q^{2}}{2(1+l)r^{2}}.
\end{equation}
The parameters $M$ and $Q$ are associated with the mass and electric charge of the central object, respectively. In the far-field ($r \rightarrow \infty$) the line element behaves as
\begin{equation}
\label{LE2}\mathrm{d}s^{2} = -\mathrm{d}t^{2}+(1+l)\mathrm{d}r^{2} + r^{2}\mathrm{d}\Omega^{2}.
\end{equation}
Therefore, the spacetime is not asymptotically flat due to the presence of the LV parameter $l$~\footnote{If we rescale the radial coordinate as $\tilde{r} = \sqrt{1+l}\, r$, the spacetime describes a geometry with a solid deficit (or excess) angle, which is mathematically identical to the asymptotic spacetime of a global monopole~\cite{BezerradeMello:2001pg}.}.

\begin{figure}[!htbp]
\begin{centering}
    \includegraphics[width=1\columnwidth]{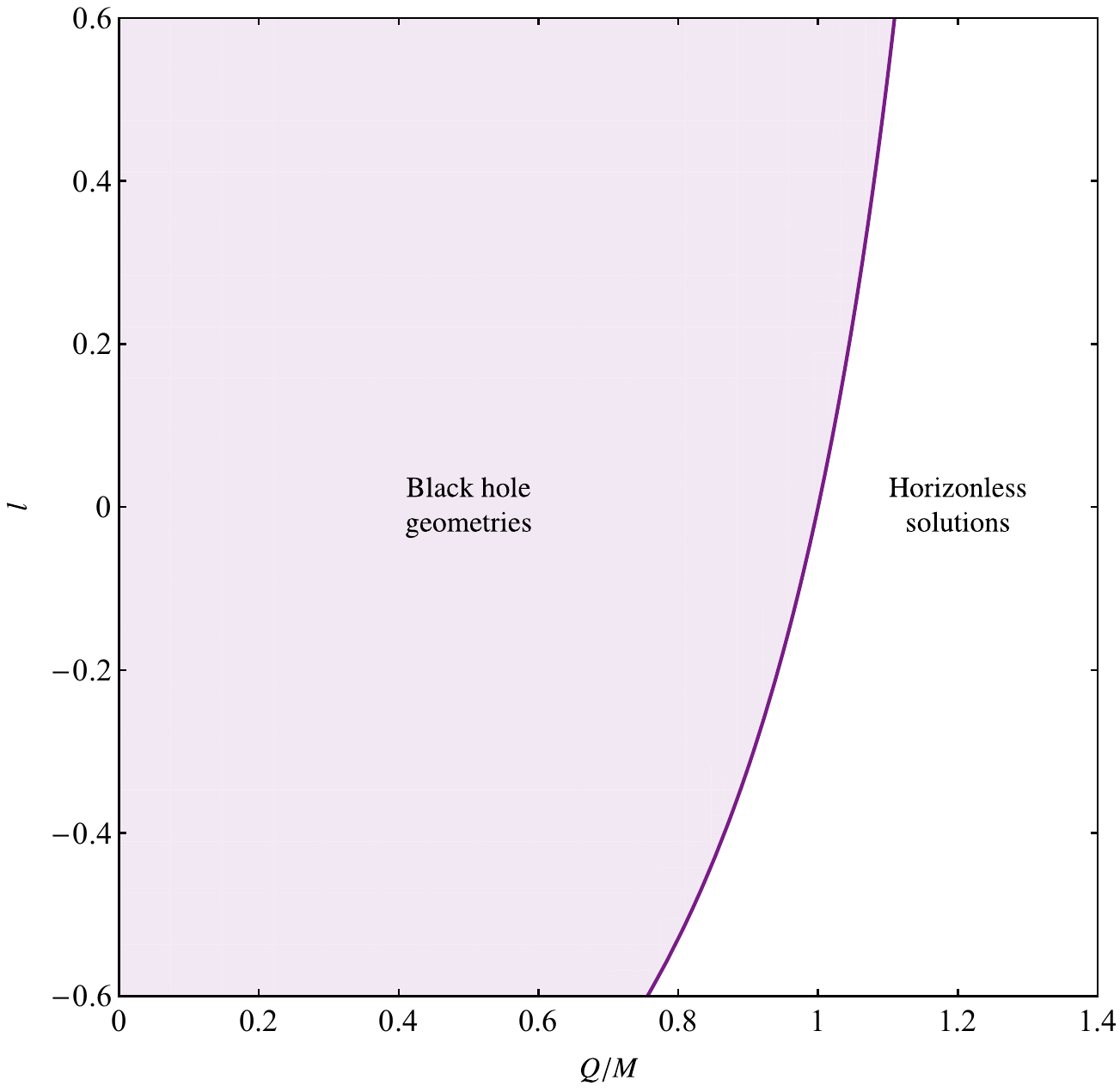}
    \caption{Existence lines of charged BHs in bumblebee gravity. The purple curve denotes the extreme charge case, given by Eq.~\eqref{extremecase}. In this figure, the extremal charge, considering $l = -0.6$ and $l = 0.6$, is $Q_{\rm{ext}} = 0.7559M$ and $Q_{\rm{ext}} = 1.1094 M$, respectively.}
    \label{bhsolutions}
\end{centering}
\end{figure}

We can obtain the horizons of the charged bumblebee BH from the positive roots of $f(r) = 0$, which leads to
\begin{equation}
\label{horizons}r_{\pm} = M \pm \sqrt{M^2-\frac{(l+2) Q^2}{2 (l+1)}},
\end{equation}
where the indices $\pm$ correspond to the event ($r_{+}$) and Cauchy ($r_-$) horizons, respectively. The extreme charge case $Q_{\rm{ext}}$ can be obtained by solving $f(r) = 0$ and $f(r)^{\prime} = 0$ simultaneously, where the prime symbol ($^\prime$) denotes differentiation with respect to the radial coordinate $r$. Thus, we find that
\begin{equation}
\label{extremecase}Q_{\rm{ext}}^{2} = \dfrac{2(1+l)}{2+l}M^{2},
\end{equation}
and the corresponding event horizon location is given by $r_{\rm{ext}} = M$. In Fig.~\ref{bhsolutions}, we display the existence lines of the charged BH in bumblebee gravity. Notice that the values of $Q/M$ depend on the parameter $l$ [cf. Eq.~\eqref{extremecase}], with $l = 0$ corresponding to the RN case. As we can see, the effect of $l$ is typically to increase the maximum value of the charge-to-mass ratio that the BH can hold. In particular, for $l < 0$ $(l >0)$, the extremal charge of the corresponding bumblebee BH is smaller (larger) than that of the RN case.

\begin{figure}[!htbp]
\begin{centering}
    \includegraphics[width=1\columnwidth]{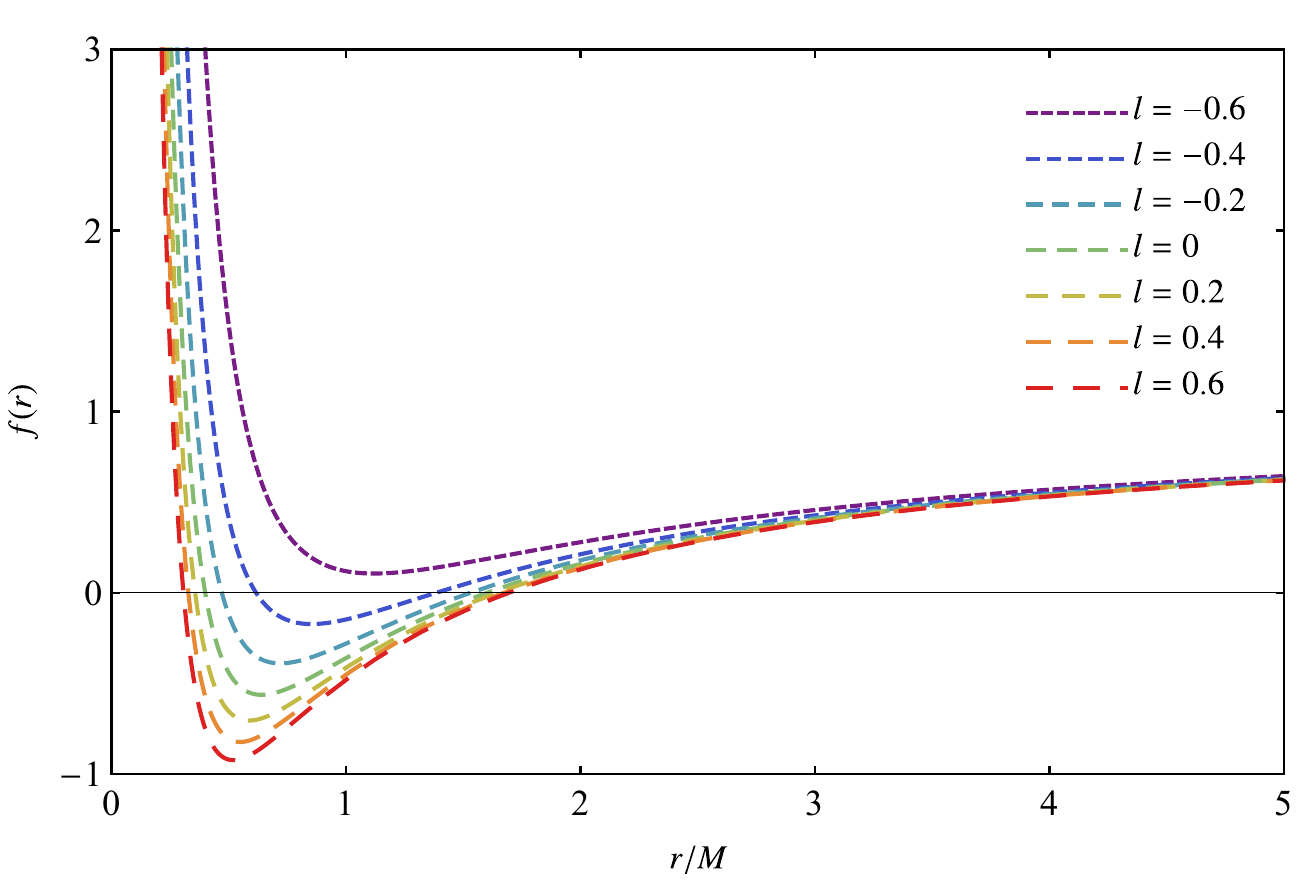}
    \caption{Metric function of the charged BH in bumblebee gravity, for distinct values of $l$, as a function of $r/M$. We fixed $Q = 0.8M$ and exhibit the corresponding case for the RN BH geometry ($l = 0$).}
    \label{mf}
\end{centering}
\end{figure}

In Fig.~\ref{mf}, we exhibit the metric function of the charged bumblebee BH geometry. We observe that BHs exist when the condition $|Q| \leq Q_{\rm{ext}}$ is satisfied. For $|Q| < Q_{\rm{ext}}$, we have up to two horizons, given by the roots of $f(r) = 0$ [cf. Eq.~\eqref{horizons}], while $|Q| = Q_{\rm{ext}}$ leads to extremal BH solutions [cf. Eq.~\eqref{extremecase}]. In turn, $|Q| > Q_{\rm{ext}}$ is related to horizonless solutions, whose scope is beyond this paper. Notice that the effect of increasing $l$ is to increase the size of the BH, i.e., the event horizon radius gets larger in accordance with Eq.~\eqref{horizons}. We also note that $l = -0.6$ corresponds to a horizonless case, as we have set $Q = 0.8M$ in accordance with Fig.~\ref{bhsolutions}. Recall that for $l = -0.6$, the extremal charge is given by $Q_{\rm{ext}} = 0.7559M$.

As discussed in Refs.~\cite{Lobo:2020ffi,Bronnikov:2012wsj}, it is possible to verify whether the spacetime is regular in the sense that all invariants constructed from the Riemann tensor and the metric tensor are finite by checking the finiteness of the Kretschmann scalar. This scalar is defined as~\cite{wald2010general}
\begin{equation}
K \equiv R_{\mu\nu\sigma\rho}R^{\mu\nu\sigma\rho},
\end{equation}
where $R_{\mu\nu\sigma\rho}$ is the Riemann tensor. For the charged bumblebee BH spacetime, we find that
\begin{align}
\nonumber K(r) = \ & \dfrac {(2 + l) Q^{2}} {(1 + l)^3r^{7}}\left[\dfrac {14 (2 + l) Q^{2}} {(1 + l) r} - 
    4 (12 M + l r) \right] + \\
\label{KS}\ & \dfrac{4\left(l^2 r^2+4 l M r+12M^{2}\right)}{(l+1)^2 r^6}.
\end{align}

\begin{figure}[!htbp]
\begin{centering}
    \includegraphics[width=\columnwidth]{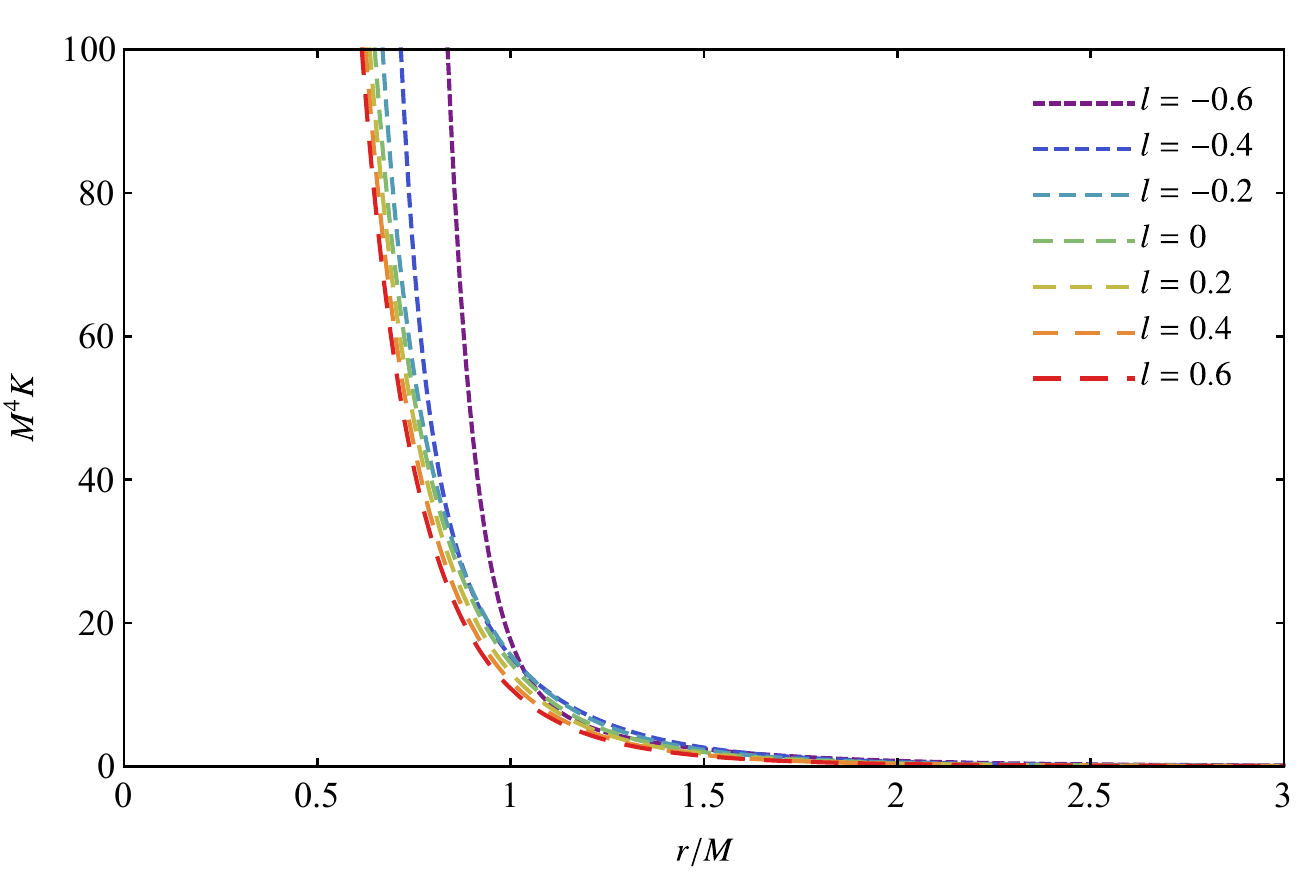}
    \caption{Kretschmann scalar invariant of the charged bumblebee BH spacetime, as a function of $r/M$, considering distinct values of $l$ with $Q = 0.7M$. Notice that $l = 0$ corresponds to the RN case.}
    \label{KSfig}
\end{centering}
\end{figure}

In Fig.~\ref{KSfig}, we display the Kretschmann scalar for the charged bumblebee BH geometry. We observe that the spacetime is singular as $r \rightarrow 0$. This means that the charged bumblebee BH spacetime inherits the same pathology as the standard BH geometries of general relativity: the classical field theory collapses at the center of the solutions.

\begin{figure}[!htbp]
\begin{centering}
    \includegraphics[width=\columnwidth]{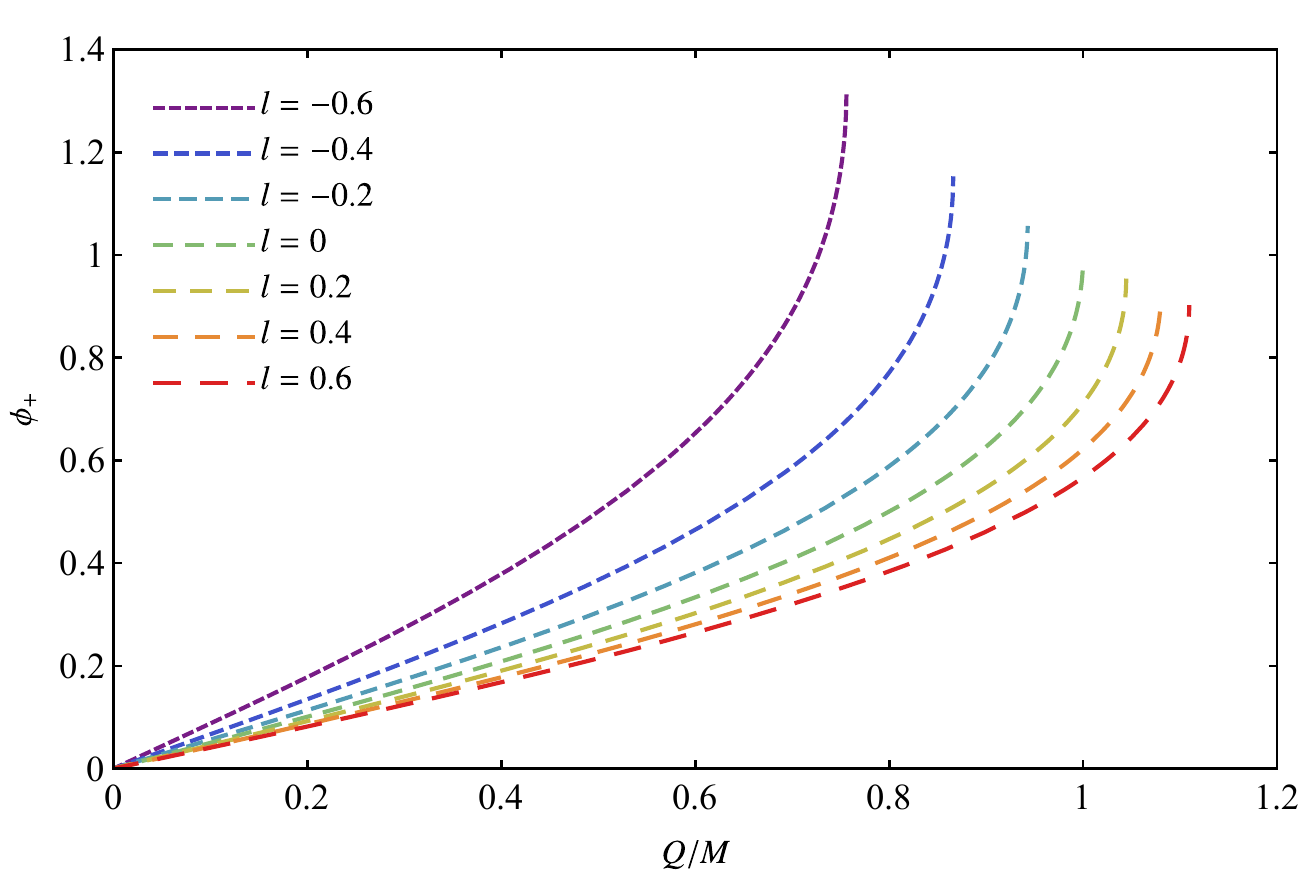}
    \caption{Radial electrostatic potential of static charged BHs in bumblebee gravity, as a function of $Q/M$, considering distinct values of $l$. The RN case is displayed in the dashed-purple curve ($l = 0$).}
    \label{epathorizon}
\end{centering}
\end{figure}

Moreover, the four-vector potential can be written as
\begin{equation}
A_{\mu} = (-\phi(r),0,0,0),
\end{equation}
where the radial electrostatic potential yields
\begin{equation}
\label{ep}\phi(r) = \dfrac{1}{r}\dfrac{(2+l)Q}{2(1+l)}.
\end{equation}
As discussed in the Introduction, the electrostatic potential evaluated at the event horizon is closely related to the threshold for superradiance in charged static BHs perturbed by complex massive fields. In Fig.~\ref{epathorizon}, we display Eq.~\eqref{ep} evaluated at the event horizon. We note that the quantity $\phi(r_+) \equiv \phi_{+}$ diminishes for fixed values of $Q/M$ as we consider higher values of $l$. Moreover, it is possible to obtain real positive values for $\phi_{+}$ with $Q/M > 1$ that satisfy $\phi_{+} < \phi_{+}^{\rm{RN}}$, as long as $l > 0$. Conversely, for $l < 0$, the real positive values for $\phi_{+}$ are characterized by $Q/M < 1$ with $\phi_{+} > \phi_{+}^{\rm{RN}}$. Therefore, when compared to the RN case, the superradiance threshold for charged BHs in bumblebee gravity is typically smaller (larger) in magnitude, resulting in a narrower (wider) superradiant frequency regime for $l > 0$ ($l < 0$).

We end this section by discussing the chosen values of $l$. From Eq.~\eqref{LE}, we observe that for any value where $l < -1$, the component $g_{rr}$ of the metric tensor flips the line element signature. In other words, $l < -1$ may be associated with nonphysical results. Later, this will also be confirmed by analyzing the propagation of charged massive scalar waves in the charged bumblebee BH spacetime [see, e.g., Eqs.~\eqref{TC0} and~\eqref{speedofthefield}]. Therefore, throughout this work, we will typically consider values of $l$ in the range $[-0.6, 2]$. These values have proven sufficient to properly investigate the influence of the LV parameter on physical and geometric quantities.

\section{Motion of charged particles}\label{sec:mcp}

The Lagrangian of a test particle with charge $\varrho$ in the equatorial plane $(\theta =\pi/2)$ of a LV charged BH is
\begin{equation}
\label{lagrangian}    \mathcal{L} = \frac{1}{2}\left(-f(r) \dot t^2 + \frac{\dot r^2}{h(r)}+r^2\dot \varphi^2\right)-\varrho \phi(r) \dot t,
\end{equation}
where the overdot denotes derivatives with respect to the proper time $\tau$ and $h(r)=f(r)/(1+l)$. The canonical momenta derived from the above Lagrangian are $\pi_\mu=\partial {\cal L}/\partial \dot x^\mu$, namely
\begin{align}
\pi_t &= -f(r) \dot t-\varrho \phi(r)=-E,\\
\pi_r&=\frac{\dot r}{h(r)},\\
\label{momenta3}\pi_\varphi &= r^2\dot\varphi = L,
\end{align}
where $E$ and $L$ are two conserved quantities conserved along the trajectory, respectively related to the energy and angular momentum of the charged particle. From a Legendre transform, one writes the Hamiltonian
\begin{align}
    \mathcal{H}&=\pi_\mu\dot x^\mu - \cal L\\
    &= \frac{1}{2}\left(-\frac{[\pi_t+\varrho\phi(r)]^2}{f(r)}+h(r) \pi_r^2+\frac{\pi_\varphi^2}{r^2}\right).
\end{align}
The Hamilton equations, namely $\dot x^\mu=\partial \cal H/\partial\pi_\mu$ and $\dot \pi_\mu = -\partial {\cal H}/\partial x^\mu$, yield the equations of motion, which in the equatorial plane read
\begin{widetext}
\begin{align}
    \dot t &= -\frac{[\pi_t+\varrho\phi(r)]}{f(r)}, \quad \dot r=h(r) \pi_r, \quad \dot \varphi=\frac{\pi_\varphi}{r^2},\\
    \dot \pi_t&=0, \quad \dot \pi_r=\frac{\pi_\varphi^2}{r^3}-\frac{h'(r)\pi_r^2}{2}+\frac{[\pi_t+\varrho\phi(r)]}{f(r)}\left[\varrho\phi'(r)-\frac{[\pi_t+\varrho\phi(r)]f'(r)}{2f(r)}\right], \quad \dot \pi_\varphi=0.
\end{align}
\end{widetext}
Moreover, since the four-velocity of massive particles is normalized ($g_{\mu\nu}\dot x^\mu \dot x^\nu=-1$), it follows that 
\begin{align*}
\dot r^2 &= U(r)\\
&= h(r)\left[\frac{[-E+\varrho\phi(r)]^2}{f(r)}-1-\frac{L^2}{r^2}\right].\numberthis
\label{motionequation}
\end{align*}
where $U(r)$ is the effective potential. 

The study of orbits is of great interest in compact object physics~\cite{chandrasekhar1998mathematical,hughes2000evolution,pani2015tidal,maggio2021extreme,teo2021spherical}. Among them, circular orbits play an important role in
astrophysics, contributing to the understanding of essential features of the dynamics of test particles around BHs. Circular orbits satisfy $\pi_r=0$ and $\dot\pi_r=0$, and consequently, along them, the energy and angular momentum obey
\begin{widetext}
\begin{align} E&=\frac{f(r)\left(2\varrho\phi(r)+\varrho r \phi'(r)+\sqrt{4f(r)-2r f'(r)+\varrho^2 r^2[\phi'(r)]^2}\right)-\varrho r \phi(r) f'(r)}{2f(r)-r f'(r)}\Bigg|_{r=r_c},\\ L&=\left.\frac{1}{r f'(r)-2f(r)}\sqrt{r^3\left(2f(r)\left[f'(r)+\varrho\phi'(r)\left(\varrho r \phi'(r)+\sqrt{4f(r)-2r f'(r)+\varrho^2 r^2[\phi'(r)]^2}\right)\right]-r[f'(r)]^2\right)}\right\vert_{r=r_c},
\end{align}
\end{widetext}
where $r_c$ is the radius of the corresponding circular orbit. The stability of these orbits can be assessed through the sign of $U''(r_c)$. Specifically, $U''(r_c)>0$ corresponds to unstable circular orbits (UCOs), $U''(r_c)<0$ corresponds to stable circular orbits (SCOs), and $U''(r_c)=0$ defines marginally stable orbits, known as the innermost stable circular orbits (ISCOs), for which $r_c=r_{\rm ISCO}$. 

\begin{figure}[h]
    \centering
    \includegraphics[width=\columnwidth]{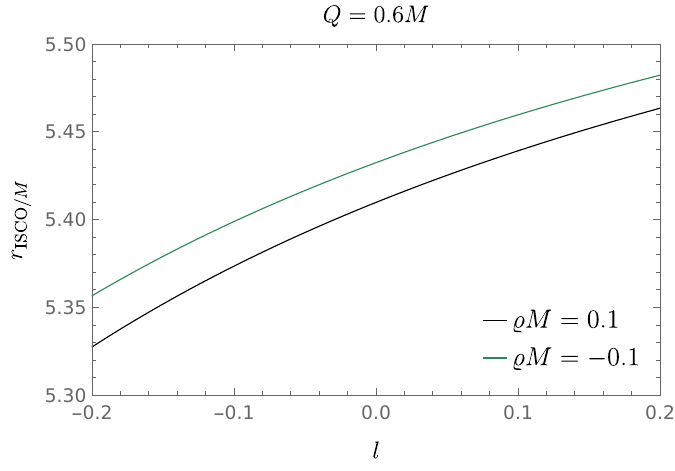}
    \caption{ISCO location for particles with electric charge $\varrho M=0.1$ and $\varrho M=-0.1$ in charged bumblebee BHs with different LV parameters and fixed charge $Q=0.6M$.}
    \label{fig:ISCO}
\end{figure}

In Fig.~\ref{fig:ISCO}, we plot the ISCO radius for particles with electric charge $\varrho M=0.1$ and $\varrho M=-0.1$ in charged bumblebee BHs with charge $Q=0.6M$ and different LV parameters. As one can see, $r_{\rm ISCO}$ increases as $l$ increases regardless of the sign of the electric charge of the particle. For an extensive analysis of the ISCO location of charged particles in charged GR BHs, we refer to Ref.~\cite{schroven2021innermost}.

Circular orbits are the simplest bound orbits one can investigate in the spacetime. These geometries, however, allow for a broad class of bound orbits. In order to investigate charged particles' bound orbits, we define a new radial coordinate $u=1/r$ and write Eq.~\eqref{motionequation} as $(\mathrm{d}u/\mathrm{d}\varphi)^2=U(1/u)/L^2$. The right-hand side of this equation is a fourth-order polynomial of $u$, namely $P(u)$, and therefore has four roots, $u_i (i=1,2,3, \text{ and }4)$, such that it can be written as
\begin{align*}
P(u)&= -\frac{(2+l)Q^2}{2(1+l)^2}u^4+\frac{2M}{1+l}u^3\\+&\left(\frac{(2+l)(-2(1+l)+(2+l)\varrho^2)Q^2}{4L^2(1+l)^3}-\frac{1}{1+l}\right)u^2\\+&\frac{(2(1+l)M-E(2+l)\varrho Q)}{L^2(1+l)^2}u+\frac{E^2-1}{L^2(1+l)}\\
&= \frac{(2+l)Q^2}{2(1+l)^2}(u_1-u)(u_2-u)(u_3-u)(u-u_4).\numberthis
\end{align*}
The polynomial $P(u)$ determines the geometry of the orbits, with the roots of $P(u)$ representing the turning points in the $r$-motion, once $\dot r=0$ at $u_i$~\cite{chandrasekhar1998mathematical}. Let us restrict our attention to the case where $u_i$ represent four real roots ordered by $u_1\geq u_2\geq u_3 \geq u_4$. Since we are looking for bound orbits, it is convenient to parametrize the roots $u_i$ in terms of parameters associated with the geometry of an ellipse, namely the eccentricity $e$ and the latus rectum $\lambda$~\cite{chandrasekhar1998mathematical}. Following Refs.~\cite{lim2024energies,chan2025periodic}, let $u_3=(1+e)/\lambda$ and $u_4=(1-e)/\lambda$ denote, respectively, the $u$-component of the periapsis and of the apoapsis. From Viète's formulas~\cite{vinberg2003course}, the roots $u_1 \equiv u_1(e,\lambda)$ and $u_2 \equiv u_2(e,\lambda)$ are related through
\begin{align}
u_1 u_2 &= \frac{2(1-E^2)(1+l)\lambda^2}{(1-e^2)(2+l)L^2Q^2},\\
u_1+u_2 &= \frac{4(1+l)M}{(2+l)Q^2}-\frac{2}{\lambda}.
\end{align}
Similarly, from Viète's formulas, one can obtain the expressions for the energy and angular momentum along the bound orbits in terms of the eccentricity and the latus rectum, which can be obtained by solving 
\begin{widetext}
\begin{align}
u_1 (u_2 + u_3+u_4) + u_2 (u_3 + u_4) + u_3 u_4 &= \frac{2(1 + l) - (2 + l) \varrho^2}{2 L^2(1 + l)} + \frac{2 (1 + l)}{(2 + l) Q^2}, \\
u_1 (u_2 u_3 +  u_2 u_4 + u_3 u_4) + u_2 u_3 u_4 &= \frac{4 (1 + l) M - 2 E (2 + l) \varrho Q}{L^2 (2 + l) Q^2}.
\end{align}
\end{widetext}
Since the expressions for $E$ and $L$ in terms of $e$ and $\lambda$ are cumbersome, we do not show them here.

To determine the period of the orbits, one has to integrate $\mathrm{d}u/\mathrm{d}\varphi=\sqrt{P(u)}$ in order to obtain $\varphi(u)$, namely
\begin{equation}
\varphi(u)=\int_{u_0}^u \frac{\mathrm{d}u'}{\sqrt{\tfrac{(2+l)Q^2}{2(1+l)^2}(u_1-u')(u_2-u')(u_3-u')(u'-u_4)}},
\end{equation}
which takes the form of an elliptic integral, with $u_0$ being the inverse of the initial radial position. Considering a periodic motion between the apoapsis and the periapsis, $1/u_4\leq r\leq 1/u_3$, during a period, the evolution of the azimuthal angle is
\begin{equation}
    \Delta\varphi_r = 2\varphi(u_3) = \dfrac{1}{\sqrt{\dfrac{(2+l)Q^2}{2(1+l)^2}}}\frac{4 K\left(\sqrt{\frac{(u_1-u_2)(u_3-u_4)}{(u_1-u_3)(u_2-u_4)}}\right)}{\sqrt{(u_1-u_3)(u_2-u_4)}},
\end{equation}
where $K(k)$ is the complete elliptic integral of the first kind with elliptic modulus $k$. One can define a parameter $\zeta$ by
\begin{equation}
\label{eq:periodic_condition}
    \zeta+1=\frac{\Delta\varphi_r}{2\pi}=\dfrac{1}{\sqrt{\dfrac{(2+l)Q^2}{2(1+l)^2}}}\frac{2 K\left(\sqrt{\frac{(u_1-u_2)(u_3-u_4)}{(u_1-u_3)(u_2-u_4)}}\right)}{\pi\sqrt{(u_1-u_3)(u_2-u_4)}}.
\end{equation}
Therefore, a periodic orbit occurs if $\Delta\varphi_r$ is a rational multiple of $2\pi$, and consequently $\zeta\in \mathbb{Q}$. This rational number is frequently written in terms of three non-negative integers $(z,w,v)$, which respectively stand for the zoom, whirl, and vertex numbers, according to the taxonomy scheme introduced in Ref.~\cite{levin2008periodic}. Specifically, the zoom number gives the number of ``leaves'' of the periodic orbit, the vertex number determines the order in which each petal is traced
out for a given $z$, and the whirl number stands for the number of laps it
executes around the BH in the time between successive leaves. In particular, $v$ and $z$ satisfy the relations
\begin{align}
    &1\leq v\leq z-1, \text{ if $z$ and $v$ are co-primes},\\
    &v=0, \text{ if $z=1$}.
\end{align}

\begin{figure}[h]
    \centering
    \includegraphics[width=\columnwidth]{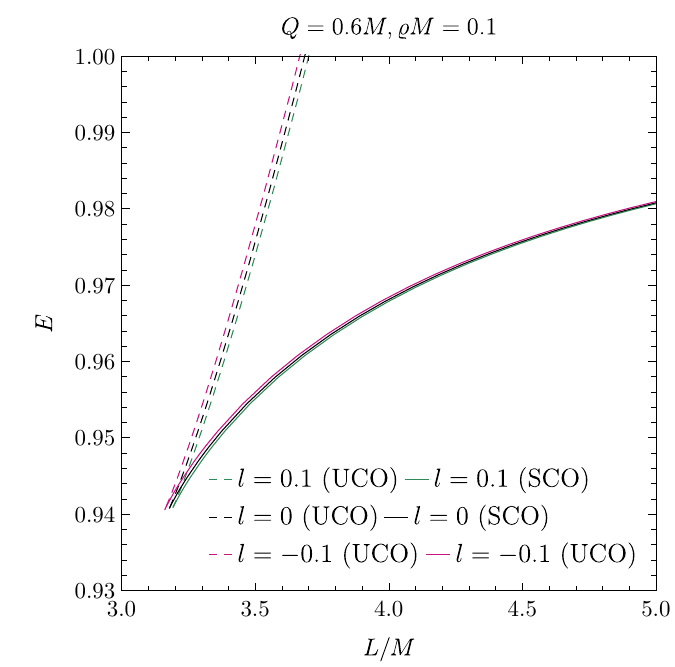}
    \caption{Values of angular momenta $L$ and energy $E$ for charged circular orbits in charged bumblebee BHs. The dashed lines stands for UCOs and the solid lines stand for SCOs. The region bounded by the two branches of circular orbits and the line $E=1$ determines the region in the $(L,E)$ space where bound orbits are allowed.}
    \label{fig:EandL}
\end{figure}

In Fig.~\ref{fig:EandL}, we show the region of the $(L,E)$ plane in which charged bound orbits exist for charged bumblebee BHs with $Q=0.6M$ and different values of $l$, for a test particle of charge $\varrho M=0.1$. The allowed domain $\mathcal{D}(Q,l;\varrho)$, bounded by the two branches of circular orbits and the line $E=1$, shifts to the left as $l$ decreases. For fixed $Q$ and $l_i\neq l_j$, the corresponding domains are not identical, $\mathcal{D}(Q,l_i;\varrho)\neq \mathcal{D}(Q,l_j;\varrho)$. Therefore, bound orbits that exist for one value of $l$ may not exist for another; in particular, orbits allowed in a charged bumblebee BH are not necessarily allowed in the RN limit $l=0$, and vice versa. In particular, one notices that values of energy and angular momentum that correspond to UCOs (SCOs) in a charged bumblebee BH with parameter $l_i$ do not correspond to any bound orbits in charged bumblebee BHs with $l_j>l_i$ ($l_j<l_i$). Conversely, values of energy and angular momentum that correspond to UCOs (SCOs) in a charged bumblebee BH with parameter $l_i$ can correspond to (non-circular) bound orbits in charged bumblebee BHs with $l_j<l_i$ ($l_j>l_i$).

\begin{figure*}
    \centering
\includegraphics[width=0.33\linewidth]{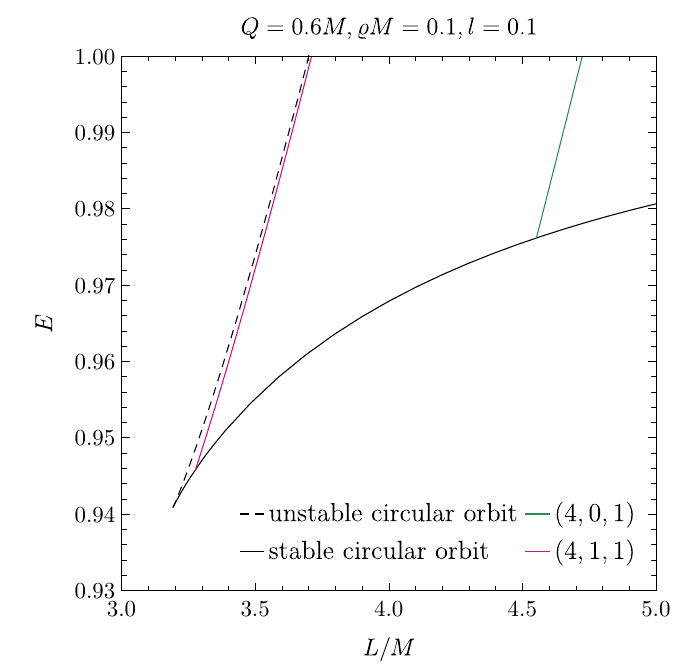}\includegraphics[width=0.33\linewidth]{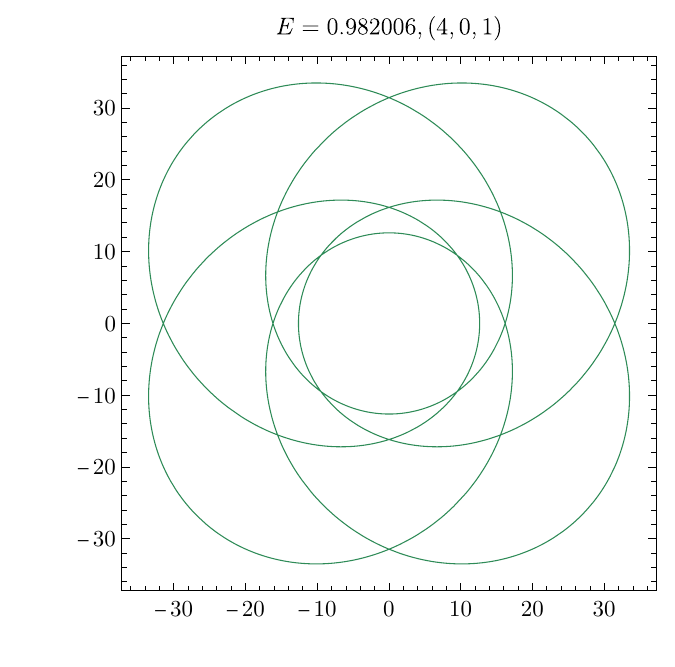}\includegraphics[width=0.33\linewidth]{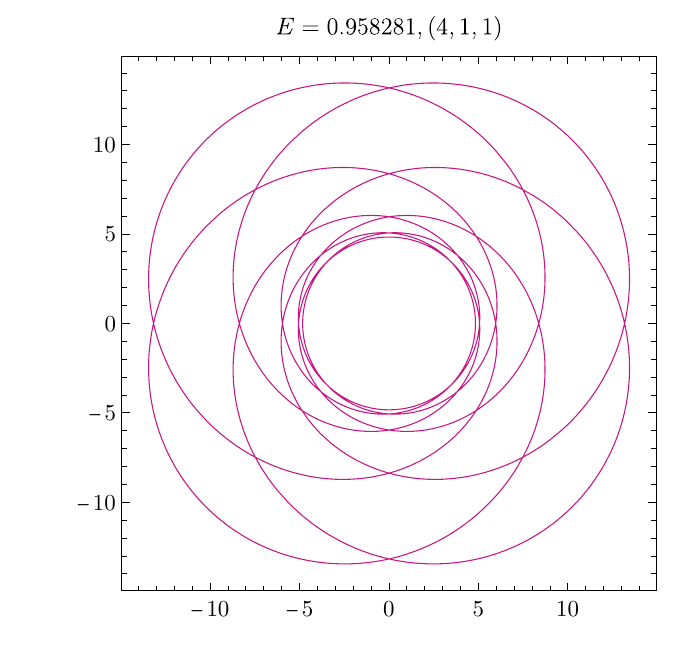}\\
    \includegraphics[width=0.33\linewidth]{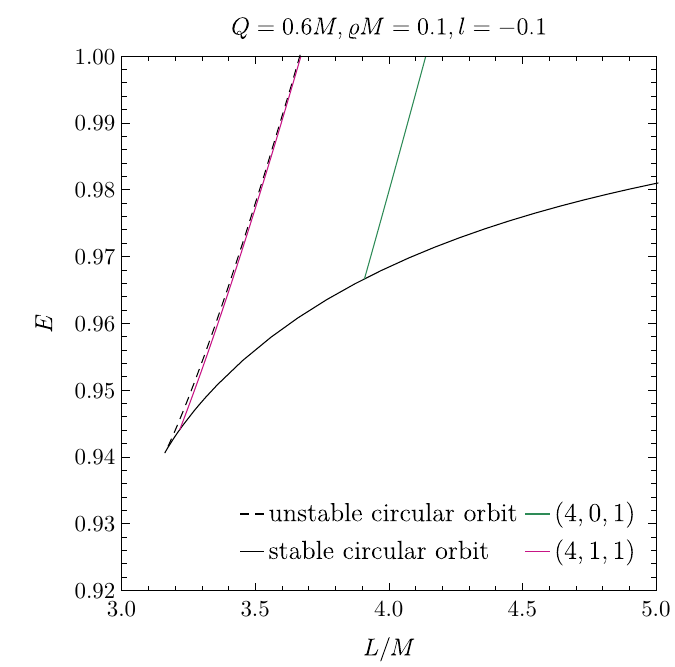}\includegraphics[width=0.33\linewidth]{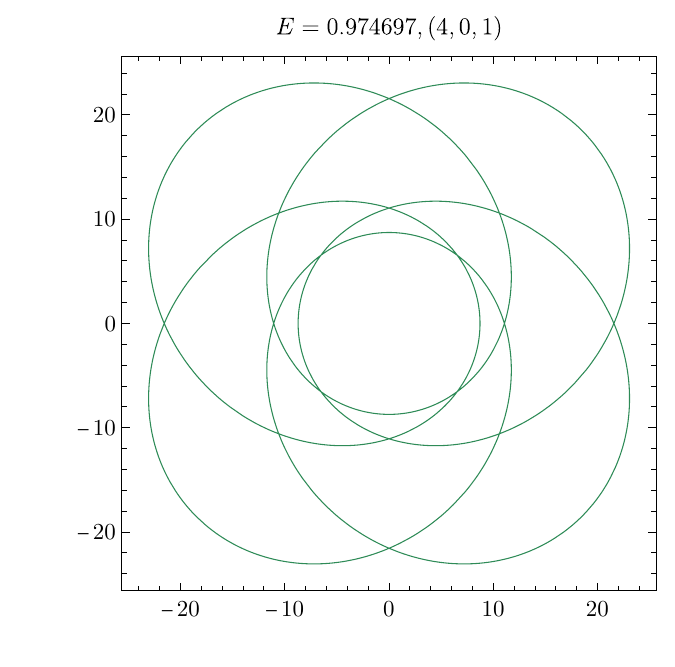}\includegraphics[width=0.33\linewidth]{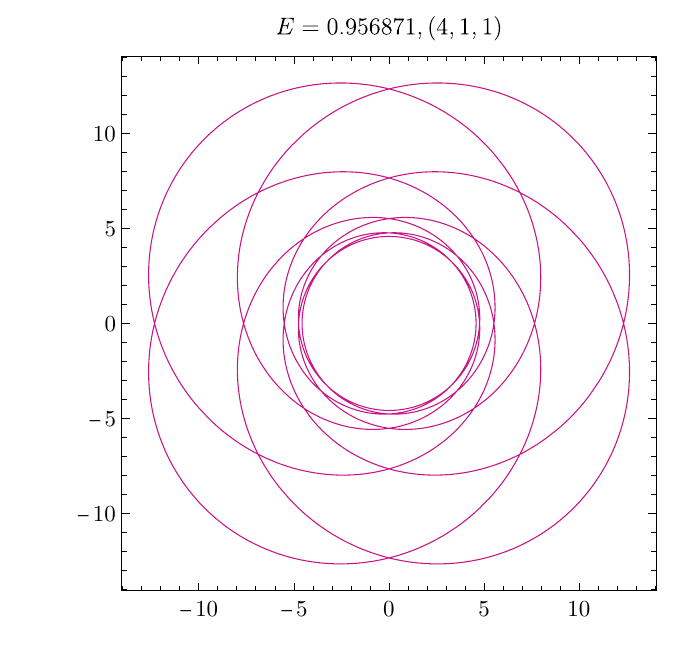}
    \caption{Left panels: (4,0,1) and (4,1,1) branches in the domains $\mathcal{D}(0.6M,0.1;0.1M)$ (top) and $\mathcal{D}(0.6M,-0.1;0.1M)$ (bottom). Middle panels: (4,0,1) periodic orbit with eccentricity $e=0.5$. Right panels: (4,1,1) periodic orbit with eccentricity $e=0.5$.}
    \label{fig:EandL_orbits}
\end{figure*}

Given a triple $(z,w,v)$ (or a rational $\zeta$), there is a curve in the $(L,E)$ space with endpoints in the SCO branch and in the line $E=1$~\cite{lim2024energies}. The endpoint on the line $E=1$ is the limit of an escaping trajectory.
These curves, called in our notation $\zeta$-branches, are parametrized by the eccentricity and latus rectum. For a periodic orbit $(z,w,v)$ with eccentricity $0< e <1$, one can solve Eq.~\eqref{eq:periodic_condition} for $\lambda$ through a root-finding method, such as Newton's method, and find the corresponding parameters $E(e,\lambda)$ and $L(e,\lambda)$ along the periodic orbits. The limit $e\to0$ corresponds to a point in the SCO branch, while the limit $e\to1$ corresponds to a point on the line $E=1$. In Fig.~\ref{fig:EandL_orbits}, we show two $\zeta$-branches, corresponding to the $(4,0,1)$ and $(4,1,1)$ periodic orbits, for charged bumblebee BHs with positive and negative values of $l$. For smaller values of $l$, the $\zeta$-branches shift to the left. This indicates that, for fixed $Q$, the same class of periodic orbits requires larger energy and angular momentum in a BH with larger $l$. In Fig.~\ref{fig:EandL_orbits}, we also exhibit, for each BH, two examples of periodic orbits with eccentricity $e=0.5$. Specifically, the panels in the middle show (4,0,1) periodic orbits, and the panels on the right show (4,1,1) periodic orbits. 
\section{Scalar wave}\label{sec:sw}
In this section, we present the wave equation associated with the propagation of charged massive scalar waves in the vicinity of static BHs. We also discuss the corresponding effective potential and the superradiance phenomenon.

\subsection{Wave equation and superradiance}\label{subsec:sup}

We are interested in investigating a scalar field $\Phi$ with mass $\mu$ and charge $q$ that is propagating in the background of the (electrically) charged bumblebee BH [cf. Eq.~\eqref{LE}]. In this context, the corresponding Klein-Gordon equation is given by
\begin{equation}
\label{KG}\left(\nabla_{\nu}-iqA_{\nu}\right)\left(\nabla^{\nu}-iqA^{\nu}\right)\Phi - \mu^{2}\Phi = 0.
\end{equation}
Using the separability of Eq.~\eqref{KG}, we can decompose $\Phi$ as
\begin{align}
\nonumber \Phi & \equiv \sum_{\ell=0}^{\infty}C_{\omega \ell} \Phi_{\omega \ell}  \\
\label{PHI0} \Phi & = \dfrac{1}{r}\sum_{\ell=0}^{\infty}C_{\omega \ell}\Psi_{\omega \ell}(r)P_{\ell}(\cos\theta)e^{-i\omega t},
\end{align}
where $\Psi_{\omega \ell}(r)$ are radial functions and $P_{\ell}(\cos\theta)$ are the Legendre polynomials. The indices $\omega$ and $\ell$ denote the frequency and the angular momentum of the scalar wave, respectively. The quantities $C_{\omega \ell}$ are constant coefficients that will be determined by the boundary conditions. By inserting Eq.~\eqref{PHI0} into Eq.~\eqref{KG}, we find the following radial equation:
\begin{equation}
\label{RE} \frac{\mathrm{d}^{2}}{\mathrm{d}r_{\star}^{2}}\Psi_{\omega \ell} - V(r)\Psi_{\omega \ell} = 0, 
\end{equation}
where $r_{\star}$ is the tortoise coordinate defined by 
\begin{equation}
\label{TC0}\mathrm{d}r_{\star} = \dfrac{\sqrt{1+l}}{f(r)}\mathrm{d}r,
\end{equation}
and the potential function $V(r)$ reads
\begin{equation}
\label{EffP} V(r) \equiv f(r)\left[\mu^{2}+\dfrac{f^{\prime}(r)}{(1+l)r}+ \dfrac{\ell(\ell+1)}{r^{2}}\right]-\left(\omega-q\phi(r)\right)^{2}.
\end{equation}
From the form of Eq.~(\ref{RE}), we notice that in regions where $V(r) < 0$, $\Phi$ is propagative, i.e., the wave is freely traveling and oscillating through space. In turn, in regions where $V(r) > 0$, $\Phi$ is evanescent, i.e., the amplitude of the wave decays exponentially rather than oscillating. 

\begin{figure}[!htbp]
\begin{centering}
    \includegraphics[width=1.0\columnwidth]{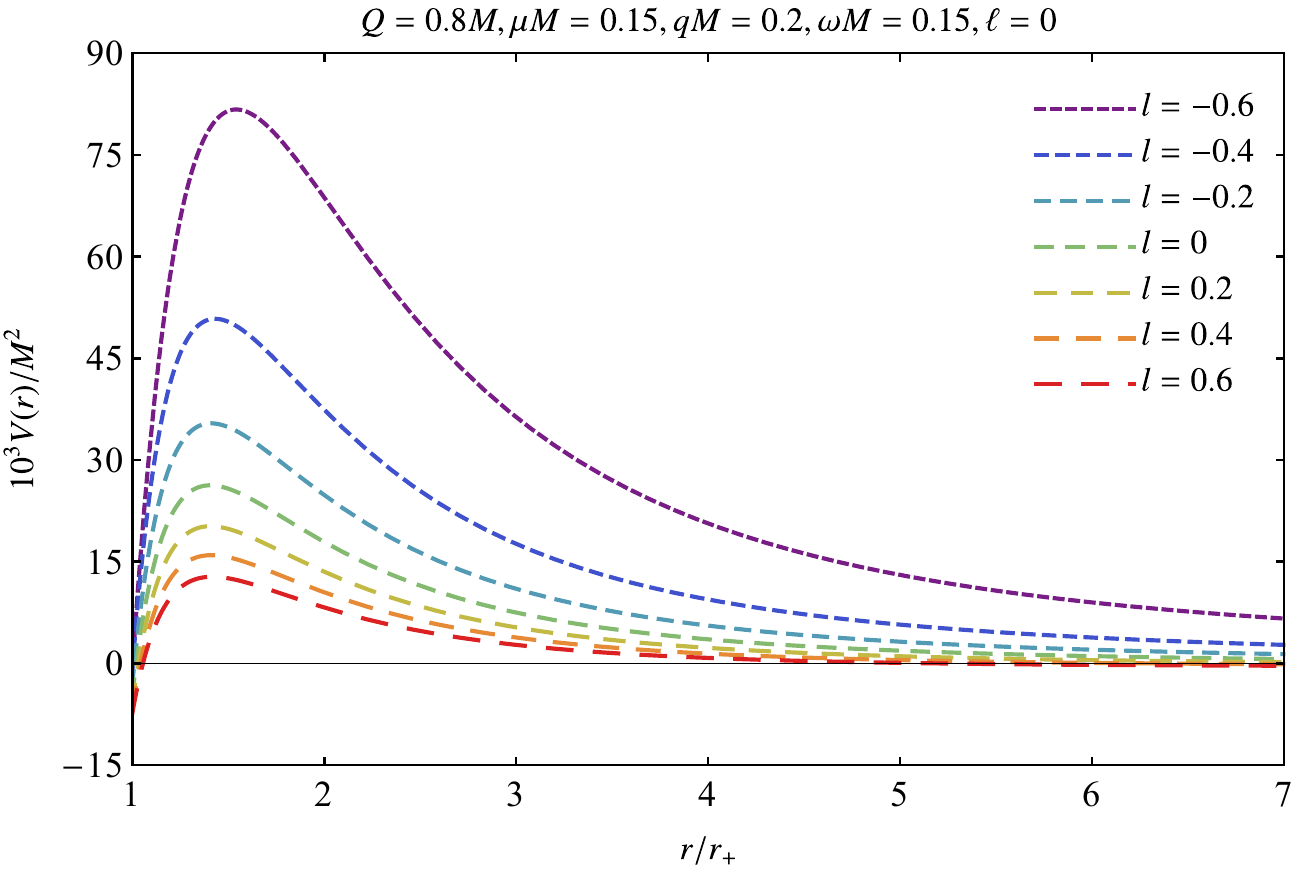}
    \includegraphics[width=1.0\columnwidth]{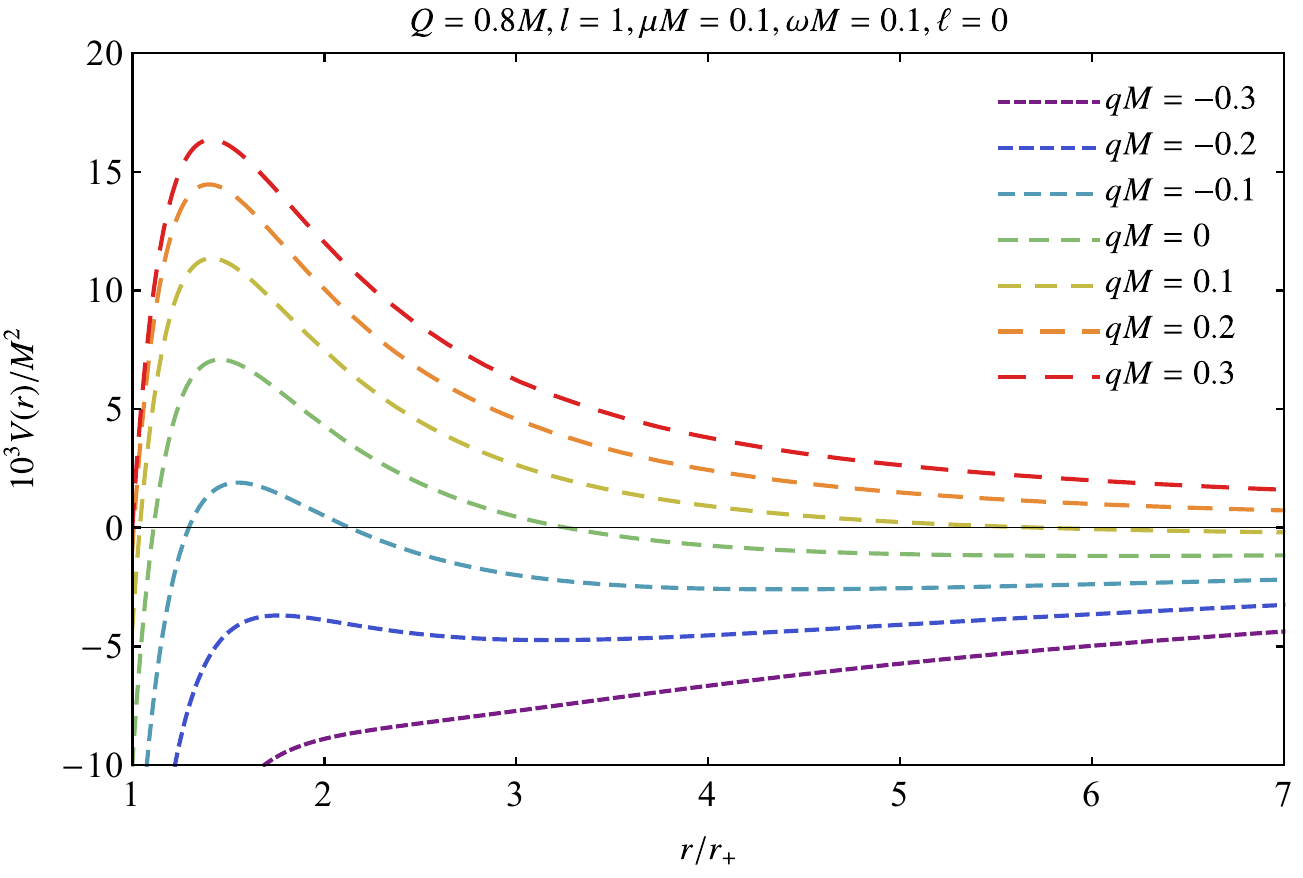}    
    \caption{Potential function of charged massive scalar waves in the background of the charged BH in bumblebee gravity, as a function of $r/r_{+}$, considering two distinct scenarios: (i) for different choices of $l$ with $Q = 0.8M$, $\mu M = \omega M = 0.15$, $qM = 0.2$, and $\ell = 0$ (top panel); and (ii) for distinct values of $qM$ with $Q = 0.8M$, $l = 1$, $\mu M = \omega M = 0.1$, and $\ell = 0$ (bottom panel).}
    \label{potentialbarrier}
\end{centering}
\end{figure}

In Fig.~\ref{potentialbarrier}, we display the function $V(r)$ for distinct values of $l$ and $qM$. Concerning the LV parameter $l$, we observe that the peak of the potential barrier decreases as we increase the values of $l$. In particular, for $l < 0$ $(l >0)$, the peak of the potential barrier of the corresponding charged bumblebee BH is higher (lower) than that of the RN case. Concerning the field charge, we see that the height of the local maximum value of $V(r)$ increases (decreases) as we consider higher values of $qM$ with $qM > 0$ ($qM <0$). This can be understood by noting that particles with the same charge sign as the BH are less absorbed than particles with the opposite charge (notice that in the bottom panel, we considered $Q = 0.8 M > 0$). In general, the local maximum of the effective potential provides a qualitative baseline for scattering: the higher the peak, the lower the chances that the wave will tunnel through the potential barrier. In turn, the lower the peak, the higher the chances that the wave will be transmitted into the BH. Moreover, for some configurations (see, e.g., the curve $qM = -0.2$ in the bottom panel), the local maximum of radial function $V(r)$ remains entirely negative outside the event horizon. 

\begin{figure}[!htbp]
\begin{centering}
    \includegraphics[width=1.0\columnwidth]{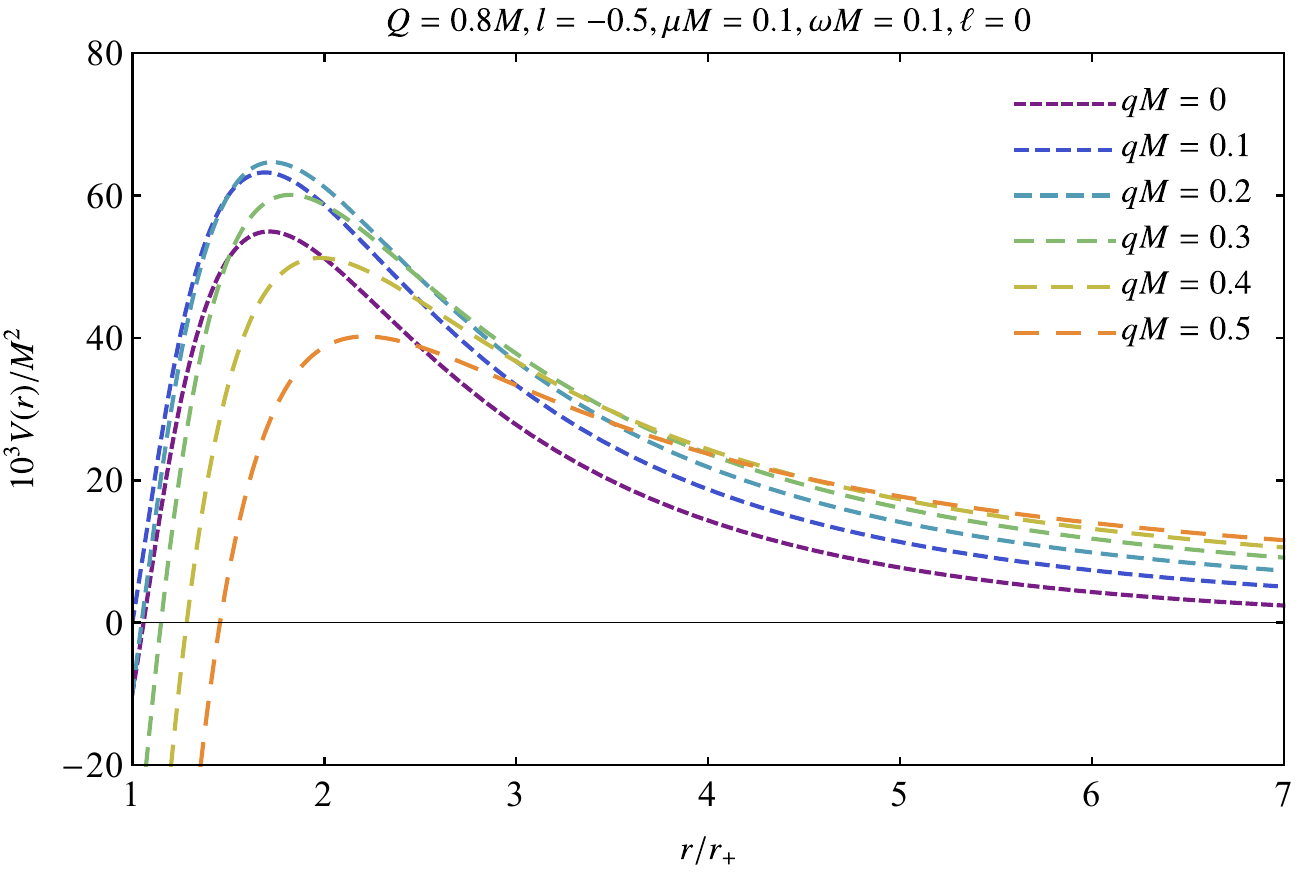}
    \caption{Potential function of charged massive scalar waves in the background of the charged BH in bumblebee gravity, as a function of $r/r_{+}$, considering distinct values of $qM$ with $Q = 0.8M$, $l = -0.5$, $\mu M = \omega M = 0.1$, and $\ell = 0$.}
    \label{effpotdifflvp2fig}
\end{centering}
\end{figure}

We also point out that the behavior of the potential barrier for the charged bumblebee BH, as we increase the BH charge-to-mass ratio or the angular momentum of the wave, is similar to what we observe for RN~\cite{Crispino:2009ki} and some regular~\cite{dePaula:2022kzz} BHs. In these cases, we observe that the peak of the potential decreases (increases) as we consider higher values of $Q/M$ ($\ell$). We notice, however, that negative values of $l$ may change the quantitative behavior of the effective potential as we vary $qM$. As shown in Fig.~\ref{effpotdifflvp2fig}, negative values of the LV parameter modify the barrier height profile. We can understand it as follows. The contribution of the electrostatic potential is encoded in the function $V(r)$ according to the term $-(\omega - q\phi(r))^2$ [cf. Eq.~\eqref{EffP}]. This term creates a sensitive competition between the repulsive superradiant contribution ($+2\omega q \phi(r)$) and the purely attractive quadratic Coulomb contribution ($-q^2 \phi(r)^2$). Consequently, for $qM > 0$ and $l < 0$, the barrier can either become larger or smaller with respect to the neutral case ($qM = 0$) depending on the value of $qM$. This leads to a non-monotonic behavior that is not observed for $l > 0$.

In the absorption/scattering problem, we are interested in waves coming from the spatial infinity towards the BH that, when interacting with a potential barrier [cf. Eq.~\eqref{EffP}], are partially transmitted into the BH and partially reflected back to spatial infinity. The boundary conditions for the scalar wave that are consistent with this scenario are given by
\begin{equation}
\label{BC}\Psi_{\omega \ell}\sim\begin{cases}
T_{\omega \ell}e^{-i \zeta r_{\star}}, & r_{\star}\rightarrow -\infty\\
e^{-i \kappa r_{\star}} + R_{\omega \ell}e^{i \kappa r_{\star}}, & r_{\star}\rightarrow+\infty
\end{cases},
\end{equation}
where $\zeta\equiv\omega-q \phi_+$, with $\phi_+ \equiv \phi(r_+)$, and $\kappa\equiv\sqrt{\omega^{2}-\mu^{2}}$. The quantities $T_{\omega \ell}$ and $R_{\omega \ell}$ are complex coefficients. For simplicity, we normalized the amplitude of the incoming scalar wave to unity. Notice also that for a propagating wave at infinity (unbounded modes), the condition $\kappa > 0$ must be satisfied. In other words, $\omega^2 > \mu^2$.

Using the Wronskian of Eq.~\eqref{RE}, one can show that
\begin{equation}
\label{CF}|R_{\omega \ell}|^{2}+\frac{\zeta}{\kappa}|T_{\omega \ell}|^{2}=1.
\end{equation}
A charged static BH can exhibit superradiance if there is a mechanism that allows the extraction of mass and charge from the BH to be feasible. In this work, this mechanism is provided by the charge of the scalar field, as discussed in Refs.~\cite{Bekenstein:1973mi,Benone:2015bst,dePaula:2024xnd}. The condition for a scalar wave to be extracted by the BH with more energy than it initially possessed requires that $|R_{\omega \ell}|^{2} > 1$. This is only possible if $\omega < \omega_{c}$, where
\begin{equation}
\label{criticalfreq} \omega_{c} = q \phi_+
\end{equation}
is the critical frequency (the threshold for superradiance). We also point out that it is useful to define the fractional gain (or loss) of energy during the scattering process. To do this, we define the amplification factor given by~\cite{BritoCardosoPani2015} 
\begin{equation}
\label{ampfactor}Z_{\omega \ell} \equiv |R_{\omega \ell}|^{2}-1 = - \frac{\zeta}{\kappa}|T_{\omega \ell}|^{2} .
\end{equation}
The positive values of $Z_{\omega \ell}$ correspond to superradiant amplification. The superradiant modes exist for frequencies satisfying $\omega < \omega_c$. In this context, the wave is amplified.

\section{Absorption of scalar waves}\label{sec:asw}

In this section, we present an expression for the total ACS obtained via the partial-wave method, as well as the corresponding analytical approximations in the low- and high-frequency regimes. For simplicity, a detailed derivation of the low-frequency approximation is presented in Appendix~\ref{appx2}. Here, we only discuss the main results concerning it. 

\subsection{Total absorption cross section}\label{subsec:tacs}

The total ACS $\sigma$ for a plane wave incident upon a spherically-symmetric BH can be expanded in partial waves. However, the charged BH obtained in bumblebee gravity is not asymptotically flat [cf. Eq.~\eqref{LE2}]. Consequently, we cannot use the same equations typically used for asymptotically flat spacetimes [see, e.g., Ref.~\cite{Benone:2014qaa}]. To obtain the correct expression for the total ACS in our case, we first note that
\begin{equation}
r_{\star} = \sqrt{1+l} \,r + C \ln\left[r\right],
\end{equation}
where $C$ is a non-elucidating constant. As we can see, the tortoise coordinate leads to an extra logarithmic phase factor in all asymptotic solutions, implying that we cannot define a ``pure'' plane wave. Notice that if we set $l = 0$, this problem persists, since we expect a logarithmic phase distortion to occur in any scattering problem with a long-range potential of $1/r$~\cite{newton2002scattering,dolan2007scattering}. Therefore, we consider that
\begin{equation}
\label{PHIACS}\Phi \sim \Phi^{\rm{P}} + \Phi^{\rm{S}},
\end{equation}
where $\Phi^{\rm{P}}$ is a monochromatic planar wave propagating along the $z$-axis, given by
\begin{equation}
\Phi^{\rm{P}} \sim e^{-i\omega t}e^{i p z},
\end{equation}
and $\Phi^{\rm{S}}$ is an outgoing scattered wave. Notice that the planar wave may be approaching the BH (positive values of $z$), as well as moving away from it (negative values of $z$).

At infinity, the momentum of the scalar wave $p$ can be obtained by noting that, in this limit, the wave behaves like a charged massive particle. Thus, in this limit, we can obtain $p$ by using the norm of the four-momentum, given by
\begin{equation}
g^{\mu\nu}\left(p_{\mu}-\varrho A_{\mu}\right)\left(p_{\nu}-\varrho A_{\nu}\right) = - m^{2},
\end{equation}
where $\varrho$ and $m$ are the charge and mass of the massive charged particle, respectively, and $p_{\mu}$ are the covariant components of the corresponding four-momentum vector. Thus, we find that
\begin{equation}
p_{r} = \pm \sqrt{\dfrac{1+l}{f(r)}\left[\dfrac{1}{f(r)}\left(E - \varrho  \phi(r)\right)^{2}-m^{2}-\dfrac{L^{2}}{r^{2}} \right]}.
\end{equation}
At infinity, one can show that
\begin{equation}
f(r) \rightarrow 1, \varrho \phi(r) \rightarrow 0, \dfrac{L^{2}}{r^{2}} \rightarrow 0.
\end{equation}
Thus, we obtain that
\begin{equation}
\label{momentum}p \equiv \lim_{r \rightarrow \infty}p_{r} = \sqrt{(1+l)(\omega^{2}-\mu^{2})} = \sqrt{1+l}\kappa.
\end{equation}

We can decompose $e^{ip z}$ as~\cite{futterman1988scattering}
\begin{equation}
\label{PHID}e^{ip z} = \sum^{\infty}_{\ell = 0}(2\ell+1)i^{\ell}j_{\ell}(p r)P_{\ell}(\cos\theta),
\end{equation}
with $j_{\ell}(\cdot)$ being the spherical Bessel function. Taking the asymptotic form of the spherical Bessel function, we get
\begin{equation}
\label{PHI2}\Phi^{\rm{P}} \sim \ \dfrac{e^{- i \omega t}}{r} \sum^{\infty}_{\ell = 0} B_{\omega \ell}\left(e^{-i \kappa r_{\star}}+e^{-i\pi(\ell+1)}e^{i \kappa r_{\star}}\right)P_{\ell}(\cos\theta),
\end{equation}
where
\begin{equation}
\label{NF}B_{\omega \ell} = \dfrac{(2\ell+1)}{2i p }e^{i\pi(\ell+1)}.
\end{equation}

Eq.~\eqref{PHI2} is the asymptotic form of the analog of a planar wave. If we choose a boundary condition such that the ingoing part of Eq.~\eqref{PHI2} resembles, in the far-field, the ingoing part of Eq.~\eqref{BC}, it follows that $C_{\omega \ell} = B_{\omega \ell}$. Thus, we find that
\begin{equation}
\label{NEWPHI}\Phi = \sum^{\infty}_{\ell = 0} B_{\omega \ell}\Phi_{\omega \ell}.
\end{equation}

The total ACS is defined as the ratio between the flux of field that goes into the BH, $|N|$, and the current of the incident planar wave, $J^{z}_{\rm{inc}}$, namely,
\begin{equation}
\label{TACS}\sigma \equiv \dfrac{|N|}{J^{z}_{\rm{inc}}}.
\end{equation}
The quantity $N$ can be written as~\cite{Benone:2014qaa}
\begin{equation}
N(r) = -\int r^{2}J_{r} \mathrm{d} \Omega,
\end{equation}
with the four-current being defined as
\begin{equation}
J^{\mu} = \dfrac{1}{2i}\left(\Phi^{\star}\nabla^{\mu}\Phi-\Phi\nabla^{\mu}\Phi^{\star} \right).
\end{equation}
One can show that
\begin{equation}
N(r) = \sum_{l = 0}^{\infty}\dfrac{\pi (2\ell+1)}{\kappa (1+l)^{3/2}}\left(1-|R_{\omega \ell}|^{2} \right),
\end{equation}
and
\begin{equation}
J^{z}_{\rm{inc}} = \dfrac{\kappa}{\sqrt{1+l}}.
\end{equation}
Therefore, the total ACS yields
\begin{equation}
\label{TACS2}\sigma = \sum_{\ell = 0}^{\infty}\sigma_{\ell},
\end{equation}
where $\sigma_{\ell}$ are the partial-wave contributions
\begin{equation}
\label{PACS}\sigma_{\ell} = \dfrac{\pi}{(1+l)\kappa^{2}}(2\ell+1)\left(1-|R_{\omega \ell}|^{2} \right),
\end{equation}
or, using Eq.~\eqref{CF}, 
\begin{equation}
\sigma_\ell = \dfrac{\pi}{(1+l)\kappa^{3}}(2\ell+1)(\omega - \omega_c)|T_{\omega \ell}|^{2}.
\end{equation}

\subsection{Low-frequency approximation}\label{subsec:lfa}

Considering $v$, we can rewrite Eq.~\eqref{eta} as
\begin{align}
\label{eta2}\eta = \dfrac{(l+2)qQ}{2(1+l)v} -\dfrac{\mu M\left(1+(1+l)v^{2}\right)}{v\sqrt{1-(1+l)v^{2}}},
\end{align}
and the low-frequency approximation~\eqref{lf} as
\begin{equation}
\label{lf2}\sigma_{\rm{lf}} = \dfrac{\pi}{\omega \sqrt{1+l}v^{2}}\left(\dfrac{4 r_{+}^{2}\zeta v \rho^{2}}{\left(1+(1+l)^{3/2}r_{+}^{2}\zeta \omega v\rho^{2}\right)^{2}+\beta^{2}\zeta^{2}} \right).
\end{equation}

Notice that the limits from the left and right of $v\rho^{2}$ are different, as $v \rightarrow 0$, defining a transition. The velocity of the transition $v_{t}$ can be found from the limit $v \rightarrow 0^{+}$, leading to
\begin{equation}
\label{vtrans}v_{t} = \pi  \left[2 \mu  M-\frac{(2+l) q Q}{1+l}\right],
\end{equation}
provided that $(1+l) ((2+l) q Q-2 (1+l) \mu  M)<0$. The case $(1+l) ((2+l) q Q-2 (1+l) \mu  M) > 0$ is discussed later. Since $l = (-1,\infty)$, we do not need to consider cases where $l$ is negative enough to reverse the sign of these inequalities.

The small velocities regime is characterized by the limit $v \lesssim v_{t}$. In this context, by taking the limit of the low-frequency approximation~\eqref{lf2} as $v \rightarrow 0$, and also considering $q Q \ll 1$ in the limit $\omega M \rightarrow \mu M$, we find that
\begin{equation}
\label{lf_000}\sigma_{\rm{lf}}^{(1)} = \dfrac{4\pi r_{+}^{2}}{(1+l)^{3/2} \mu  v}\left[(1+l)\mu-\dfrac{(2+l)qQ}{2r_{+}}\right]\rho ^2.
\end{equation}
By inserting Eq.~\eqref{rho} into Eq.~\eqref{lf_000} while taking Eq.~\eqref{eta2} into account, and then taking the limit $v \rightarrow 0$, we can show that
\begin{align}
\nonumber \sigma_{\rm{lf}}^{(1)} = \ & \dfrac{2 \pi A_{+}}{(1+l)^{5/2} \mu  v^2} \left[(1+l) \mu  M -\dfrac{(2+l)qQ}{2}\right]\times\\
\label{lf_1}& \left[(1+l) \mu-\dfrac{(2+l) q Q}{2 r_{+}} \right],
\end{align}
where $A_{+} = 4\pi r_{+}^{2}$ is the BH area.

Now, let us focus on the case $v \gtrsim v_{t}$. If we consider small values of $\omega$ and $\mu$, and neglect second order contributions of $qQ$, then $\rho \approx 1$ in the low-frequency limit. Thus, we get
\begin{equation}
\label{lf_2}\sigma_{\rm{lf}}^{(2)} = \dfrac{A_{+}}{(1+l)^{3/2} \omega  v}\left[(1+l) \omega -\dfrac{(2+l) q Q}{2 r_{+}}\right].
\end{equation}

Notice that, so far, we have discussed the case $(1+l) ((2+l) q Q-2 (1+l) \mu  M) < 0$. In what follows, we focus on the case $(1+l) ((2+l) q Q-2 (1+l) \mu  M) > 0$.  When this condition is fulfilled, the limit of Eq.~\eqref{lf2} as $\omega M \rightarrow \mu M,$ leads to
\begin{equation}
\label{lf_mM}\lim_{\omega \rightarrow \mu}\sigma_{\rm{lf}} = 0.
\end{equation}
Recall that as $\mu M \rightarrow \omega M$, $v \rightarrow 0$. In Sec.~\ref{sec:mr}, we compare our numerical results with the analytical approximations for the total ACS in the (very) low-frequency regime. We also point out that the main equations presented in this section, as well as in Appendix~\ref{appx2}, reduce to those of the RN case for $l = 0$. The neutral-massive and neutral-massless counterparts are obtained by setting $q = 0$ and $q = \mu = 0$, respectively.

\subsection{High-frequency approximation}

In the high-frequency regime, the propagation of the charged massive scalar wave can be treated in analogy to the motion of charged massive particles. Therefore,  the characteristics of the scalar waves can be associated with the trajectories of charged particles subject to the Lorentz force. The motion of charged particles was studied in Sec.~\ref{sec:mcp}. Here, we simply rewrite the main equations in a form that is more convenient for comparing them with the numerical results.

Using Eqs.~\eqref{lagrangian}-\eqref{momenta3}, we can rewrite the equation of motion [cf. Eq.~\eqref{motionequation}] of massive charged particles as
\begin{equation}
\label{Kr}\mathcal{K}(r) =\frac{h(r)}{f(r)}\left(\dfrac{E - \varrho\phi(r)}{L}\right)^2 - h(r)\left( \frac{m^2}{L^2} + \frac{1}{r^2} \right),
\end{equation}
where $\mathcal{K}(r) \equiv (m^{2}/L^{2})\dot{r}^{2}$. The impact parameter is defined as the distance between the asymptotic trajectory of the particle and the head-on collision line, namely, $b \equiv L/p$~\cite{newton2002scattering}. Notice that as the charged bumblebee BH spacetime is not asymptotically flat, the asymptotic momentum of the particle $p$ inherits a factor associated with the non-flat asymptotic behavior of the geometry [cf. Eq.~\eqref{momentum}]. Therefore, we write the impact parameter as 
\begin{equation}
\label{impactparameter}b = \dfrac{L}{(1+l)vE}.
\end{equation}
Thus, using Eqs.~\eqref{Kr},~\eqref{impactparameter}, and~\eqref{speedofthefield}, we find that
\begin{widetext}
\begin{equation}
{\mathcal{K}}(r) = \frac{1}{(1+l)^{3}b^2 v^2} \left( 1 - \frac{\sqrt{1 - (1+l)v^2}}{m} \varrho\phi(r) \right)^2 - \frac{f(r)}{1+l}\left( \frac{1 - (1+l)v^2}{(1+l)^{2}b^2 v^2} + \frac{1}{r^2} \right),
\end{equation}
\end{widetext}
for the charged bumblebee BH spacetime. The high-frequency absorption cross section is given by~\cite{wald2010general}
\begin{equation}
\label{hf}\sigma_{\rm{hf}} = \pi b_{c}^{2},
\end{equation}
where $b_{c}$ is the critical impact parameter, which is obtained by solving ${\mathcal{K}}(r_{c}) = 0$ and ${\mathcal{K}}^{\prime}(r=r_{c}) = 0$ simultaneously. These conditions lead to equations that are not elucidative for the critical impact parameter and the corresponding orbital radius, and they depend on the BH parameters. Therefore, we have chosen not to present these equations here. We emphasize, however, that in the limit where $l = 0$, we obtain the same equations presented in Ref.~\cite{dePaula:2024xnd}. The high-frequency approximation derived for the ACS obtained here can be used to predict the general behavior of the total ACS for scalar waves. In Sec.~\ref{subsec:nm}, we compare Eq.~\eqref{hf} with our numerical results.

\section{Main results}\label{sec:mr}

In this section, we present our main results concerning the absorption and superradiance of charged massive scalar waves in the background of the charged bumblebee BH. For simplicity, we divide our main results into two sections: (i) Numerical method; and (ii) superradiance of charged bumblebee BHs.

\subsection{Numerical method}\label{subsec:nm}

Our goal is to find the complex coefficients $R_{\omega \ell}$, as they dictate the behavior of the amplification factors [cf. Eq.~\eqref{ampfactor}] and total ACS [cf. Eq.~\eqref{TACS2}]. To do this, we solve Eq.~\eqref{RE} numerically from near the event horizon, i.e., $r_{\rm{ini}} = 1.0001 r_{+}$, to far away from the BH, i.e., $r_{\rm{inf}} = 10^{3}M$, using the stiffness-switching integration method~\cite{press2007numerical}. Then we match the numerical solutions of Eq.~\eqref{RE} with the appropriate boundary conditions, given by Eq.~\eqref{BC}. The linear system relating the radial function and its radial derivative at the numerical infinity with the complex coefficients can be written as
\begin{align}
\label{LS}\begin{bmatrix}
\Psi_{\omega \ell}(r) \\ 
\Psi_{\omega \ell}^{\prime}(r)
\end{bmatrix} 
=
\begin{bmatrix}
e^{-i\kappa r_{\star}} & e^{i\kappa r_{\star}} \\ 
\left(e^{-i\kappa r_{\star}}\right)^{\prime} & \left(e^{i\kappa r_{\star}}\right)^{\prime}
\end{bmatrix}\cdot
\begin{bmatrix}
1/T_{\omega \ell} \\ 
R_{\omega \ell}/T_{\omega \ell}
 \end{bmatrix}.
\end{align}
For simplicity, we normalize the coefficients of Eq.~\eqref{BC} by $T_{\omega \ell}$. Moreover, the waveform at infinity is obtained by setting
\begin{equation}
\label{psinum}\Psi_{\omega \ell} = r^{\beta_{0}} e^{i\kappa r_{\star}}g(r),
\end{equation}
where $\beta_{0}$ is obtained by solving Eq.~\eqref{RE} at infinity, with
\begin{equation}
g(r) = \sum_{i}^{n}\dfrac{g_{i}}{r^{i}}.
\end{equation}
The coefficients $g_{i}$ are determined by imposing that Eq.~\eqref{psinum} satisfies Eq.~\eqref{RE}, and the index $n$ governs the order of the asymptotic expansion. For our purposes, we set $n = 5$.

To solve the ordinary differential equations numerically, we need to impose two initial conditions. Thus, we fix the field and its radial derivative near the event horizon as
\begin{subequations}
\begin{align}
\label{bc1}\Psi_{\omega \ell}(r_{\rm{ini}}) & =  1, \\ 
\label{bc2}\Psi_{\omega \ell}^{\prime}(r_{\rm{ini}}) & = -\dfrac{i\sqrt{1+l}\zeta}{f(r_{\rm{ini}})}.
\end{align}
\end{subequations}
The numerical values for $R_{\omega l}$ are obtained by solving the linear system given by Eq.~\eqref{LS} with the aforementioned initial conditions. We also point out that the oscillatory pattern of the total ACS is associated with the partial-wave contributions [cf. Eq.~\eqref{PACS}]. We have chosen, in general, to perform the summation in Eq.~\eqref{TACS2} up to $\ell = 10$. Moreover, in the graphical analyzes, we normalize the wave frequency by the wave mass. We consider values of LV parameter in the range $l \in [-0.4, 2]$. High values of $l$, specifically $l \geq 1$, are typically chosen to improve the graphical display of the results, but, in general, the values of $l$ are expected to be small~\cite{Cordeiro:2025eox}.

\begin{figure}[!htbp]
\begin{centering}
    \includegraphics[width=1.0\columnwidth]{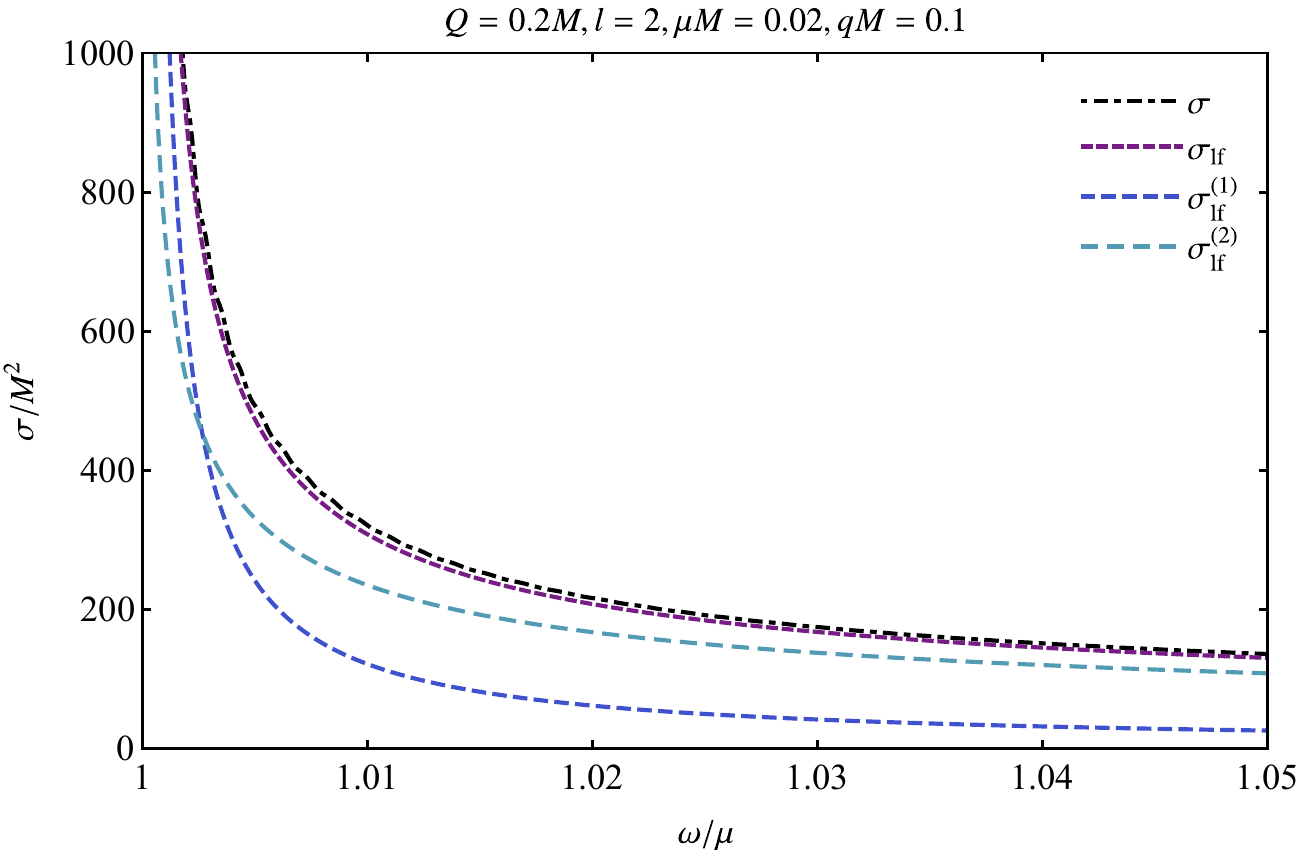}   \includegraphics[width=1.0\columnwidth]{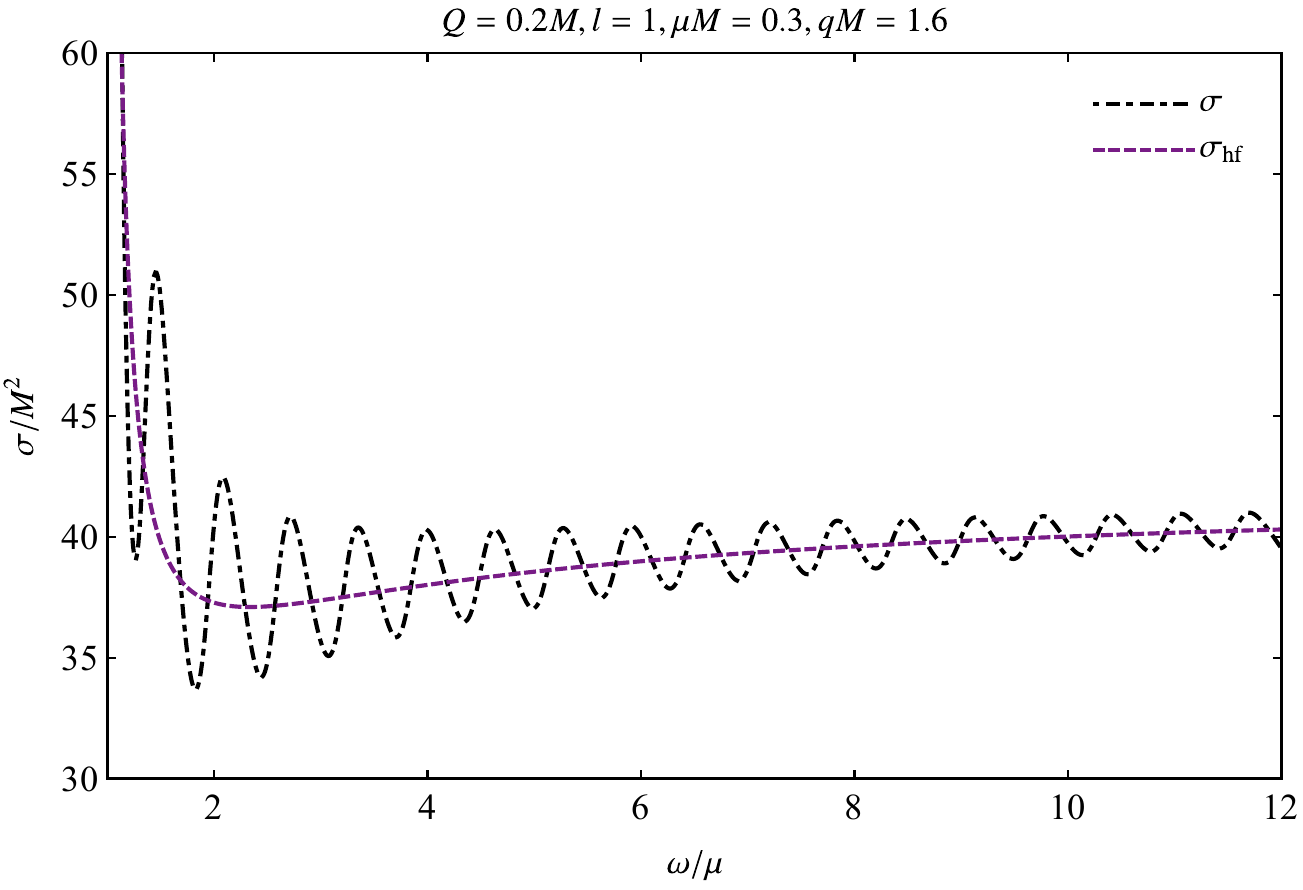}
    \caption{Comparison between the total ACS computed numerically with the analytical approximations in the low- (top panel) and high- (bottom panel) frequency regimes, as a function of $\omega/\mu$. In the top panel, we set $Q = 0.2M$, $l = 2$, $\mu M = 0.02$, and $qM = 0.1$, while, in the bottom panel, we set $Q = 0.2M$, $l = 1$, $\mu M = 0.3$, and $qM = 1.6$. Notice that the transition associated with the scalar wave velocity in the top panel occurs at $\omega \approx 1.0026 \mu$.}
    \label{landhfapprox}
\end{centering}
\end{figure}

In Fig.~\ref{landhfapprox}, we compare our numerical results with the analytical approximations in the low- and high-frequency regimes. As we can see, the total ACS oscillates around the high-frequency approximation given by Eq.~\eqref{hf}. The oscillatory profile is associated with the interference of the scalar waves propagating in the vicinity of the BH. We also observe that the numerical results are consistent with the low-frequency approximation given by Eq.~\eqref{lf2}. In particular, we clearly observe the transition velocity that splits the analytical formula given by ~\eqref{lf2} into the approximations given by Eqs.~\eqref{lf_1} and~\eqref{lf_2} obtained for $v \lesssim v_{t}$ and $v > v_t$, respectively. 

In the top panel of Fig.~\ref{landhfapprox}, the transition velocity occurs at $\omega \approx 1.0026 \mu$. We emphasize that the transition velocity was first observed for the RN case~\cite{Benone:2017xmg}, considering neutral massive scalar waves, but it was not properly discussed for charged massive scalar waves. Therefore, our analytical approximations for the low-frequency regime generalize the expression provided in Ref.~\cite{Benone:2015bst}, which considers charged massive scalar waves but covers only the branch $v > v_{t}$. See, e.g., Eq. (15) of Ref.~\cite{Benone:2015bst} and compare it with our Eq.~\eqref{lf_2} for $l = 0$. Moreover, the agreement between our numerical results and the analytical approximations, within their respective limits, provides a consistency check for our results. 

\subsection{Superradiance of charged bumblebee BHs}\label{subsec:scbbhs}

\begin{figure}[!htbp]
\begin{centering}
    \includegraphics[width=1.0\columnwidth]{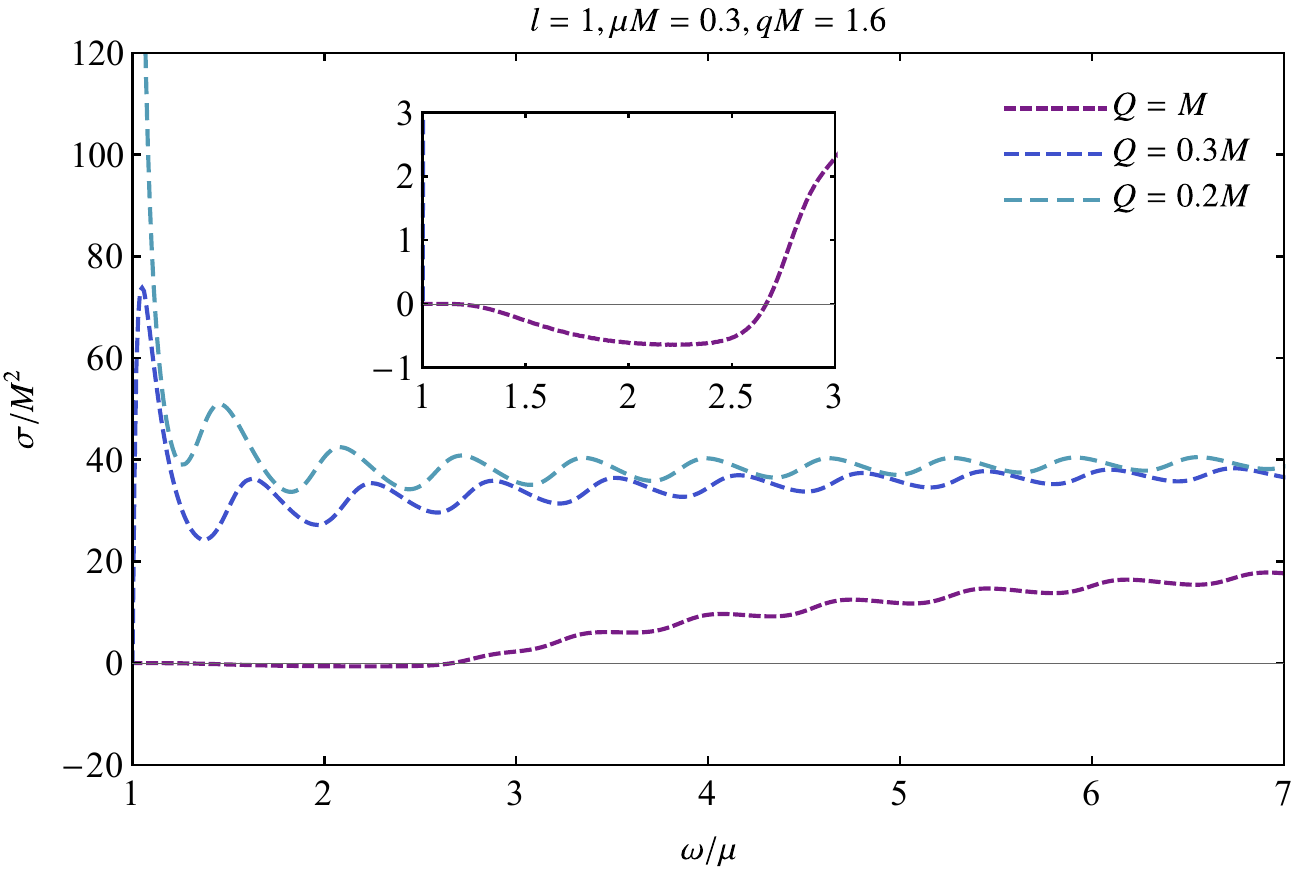}
    \includegraphics[width=1.0\columnwidth]{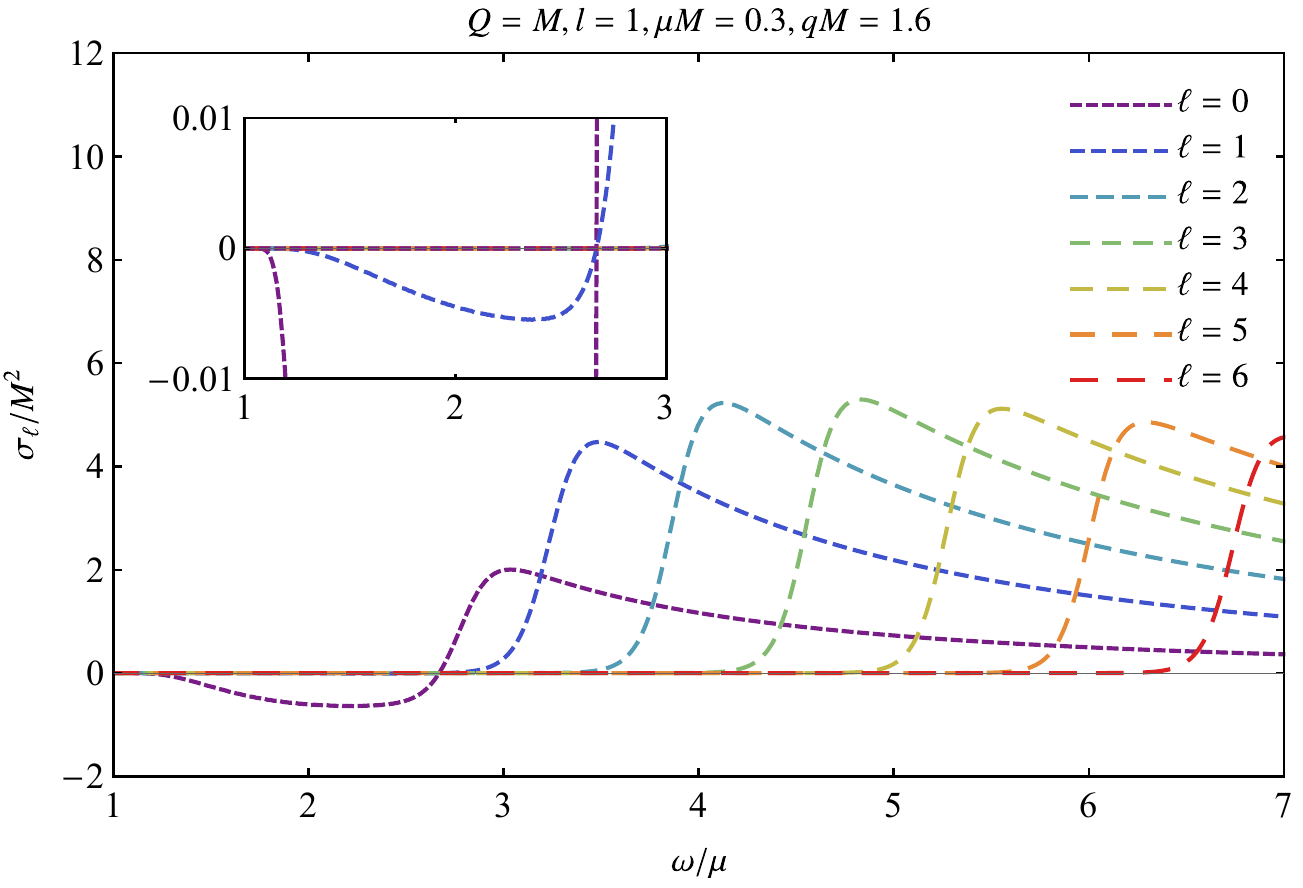}
    \caption{Top panel: Total ACS of the scalar waves in the background of the charged bumblebee BH, as a function of $\omega / \mu$, for distinct values of $Q/M$ and considering $l = 1$, $\mu M = 0.3$, and $q M = 1.6$. Bottom panel: Partial ACSs of the scalar waves in the background of the charged bumblebee BH, as functions of $\omega / \mu$, for distinct values of $\ell$ and considering $Q = M$, $l = 1$, $\mu M = 0.3$, and $q M = 1.6$. The insets in both panels zoom the superradiant scattering scenario.}
    \label{tacsdiffcrosssections}
\end{centering}
\end{figure}

According to the absorption parameter space of the charged bumblebee BH geometry (see Appendix~\ref{appx}), we might have: (i) unbounded absorption; (ii) bounded absorption; and (iii) bounded superradiance. These possibilities are shown in Fig.~\ref{tacsdiffcrosssections} (top panel). We also exhibit the corresponding partial ACSs for the case where we have bounded superradiance (bottom panel). As we can see, during the superradiant scattering process, namely for $\mu < \omega \lesssim 2.66 \mu$, the total ACS is negative. We also observe that the modes $\ell \geq 1$ contribute to the low-frequency regime of the total ACS, in contrast to the neutral massless case, where the low-frequency regime is governed solely by the fundamental mode $\ell =0$~\cite{Paula:2020yfr}. In particular, modes $\ell \geq 1$ can also contribute to the superradiant scattering, as indicated in the inset on the bottom panel.

While Fig.~\ref{tacsdiffcrosssections} demonstrates that higher angular momentum modes ($\ell \geq 1$) can carry a non-zero flux in the low-frequency regime for charged massive scalar waves, the amplitudes of these higher-order modes are heavily suppressed by the centrifugal barrier. This helps us to understand why our analytical approximation for the low-frequency regime, which is derived under the assumption $\ell = 0$ (see Appendix~\ref{appx2}), shows excellent agreement with the numerical results (see, e.g., Sec.~\ref{subsec:nm} and Fig.~\ref{landhfapprox}) in the \textit{very} low-frequency regime. 

\begin{figure}[!htbp]
\begin{centering}
    \includegraphics[width=1.0\columnwidth]{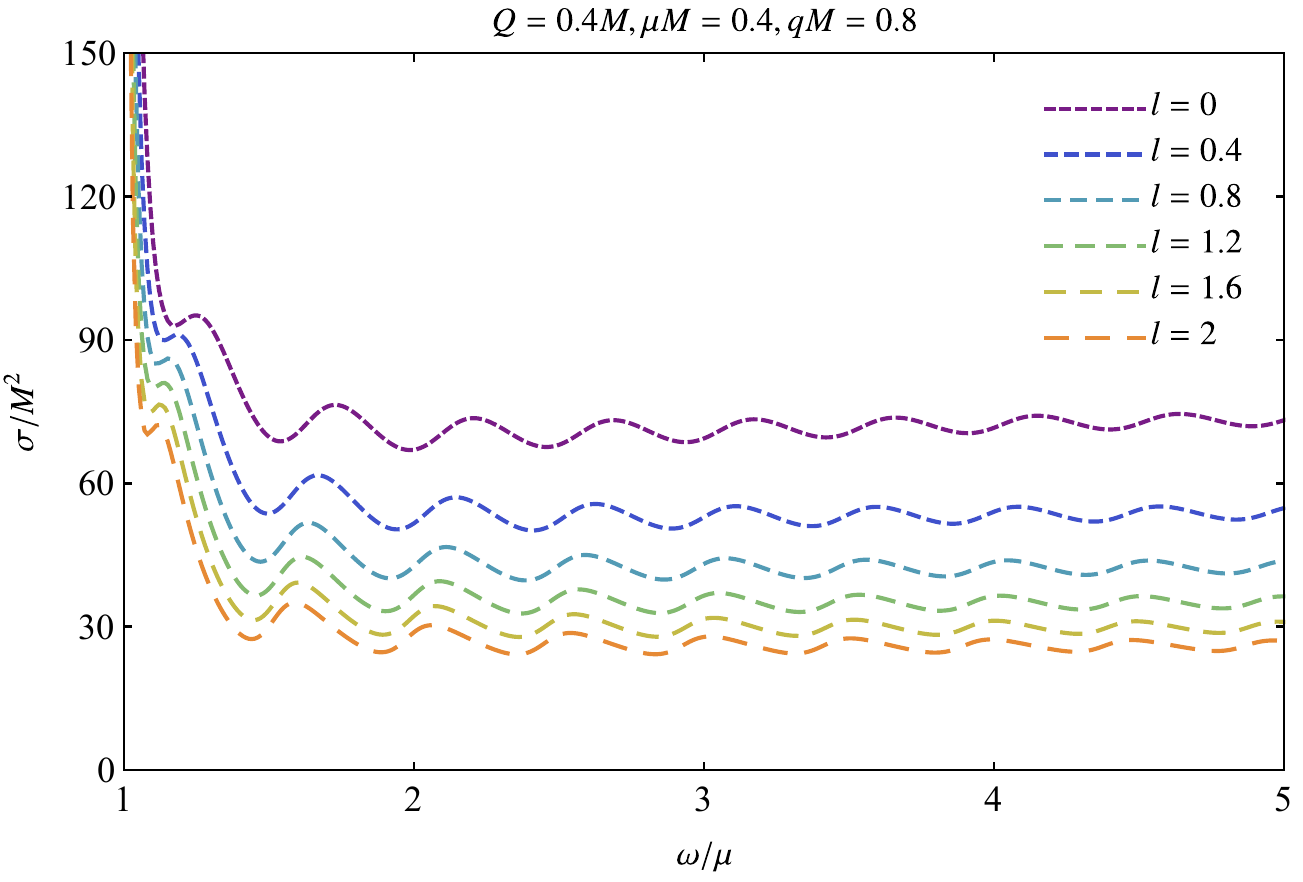}    \includegraphics[width=1.0\columnwidth]{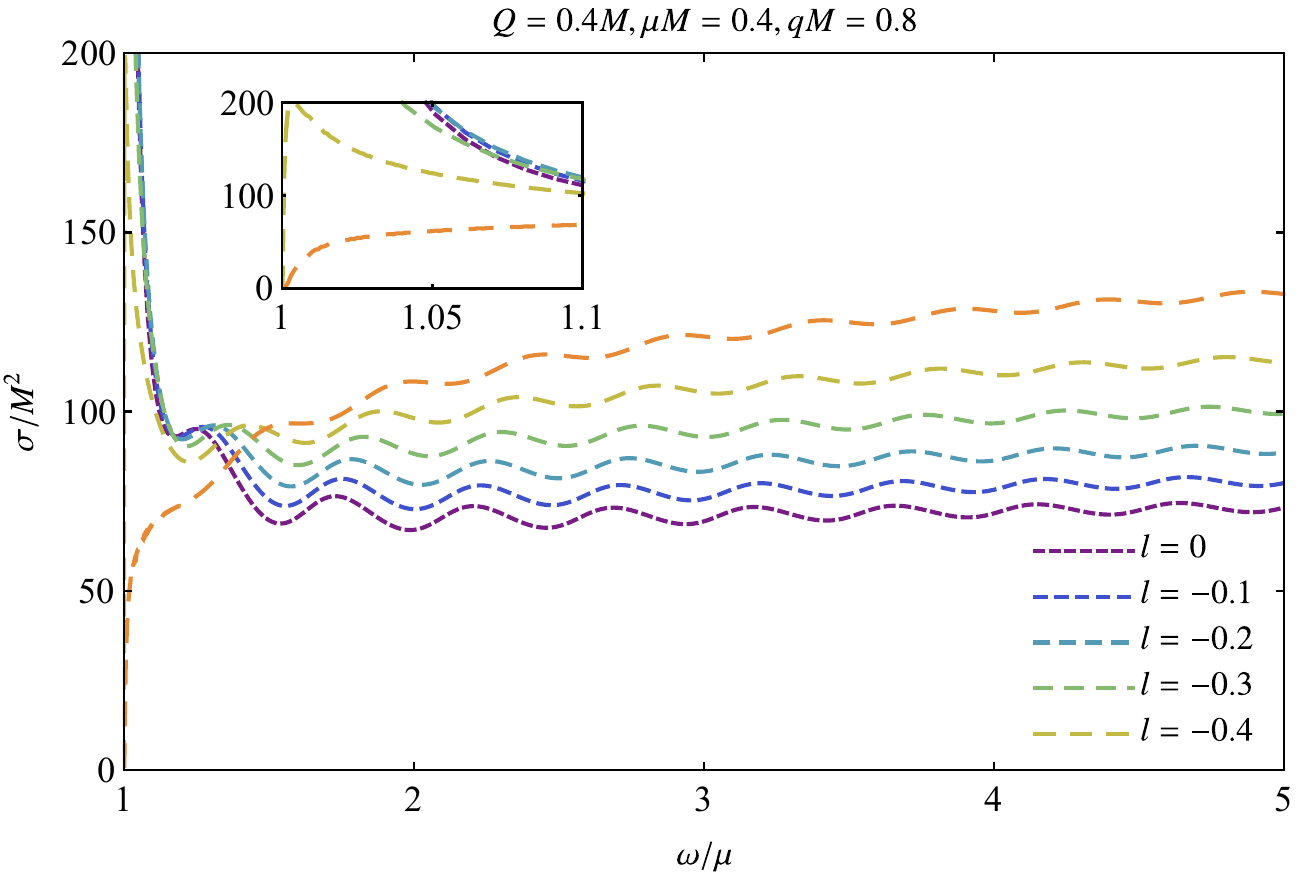}
    \caption{Total ACS of the scalar waves in the background of the charged bumblebee BH, as a function of $\omega / \mu$, for distinct values of $l$ and considering $Q = 0.4M$, $\mu M = 0.4$, and $q M = 0.8$. In the top (bottom) panel, we consider positive (negative) values for $l$.}
    \label{tacsdiffl}
\end{centering}
\end{figure}

In Fig.~\ref{tacsdiffl}, we present the total ACS of the charged bumblebee BH, considering distinct values of $l$. We observe that the total ACS decreases as we consider higher values of $l$. This behavior appears to be inconsistent with the behavior of the potential barrier (see, for example, the top panel of Fig.~\ref{potentialbarrier}), which decreases as $l$ increases; thus, one would expect the corresponding total ACSs to increase. Notice that the expression for the total ACS of the charged bumblebee BH spacetime inherits a kinematic factor of $1/(1+l)$ due to the non-flat asymptotic behavior of the spacetime [cf. Eq.~\eqref{PACS}]. While the lowered (raised) potential barrier increases the absorption (reflection) probability of the wave [cf. Eq.~\eqref{CF}], this behavior is overpowered by the global suppression of the $1/(1+l)$ geometric prefactor as $l$ increases. Therefore, although the potential barrier is useful for predicting certain characteristics of the total ACS, it must be analyzed carefully, particularly when investigating physical observables, such as the total ACS, in spacetimes that are not asymptotically flat. We also observe that, in the bottom panel of Fig.~\ref{tacsdiffl}, we have bounded absorption for $l = -0.3$ and $l = - 0.4$, as expected since for these values of $l$, the chosen BH and scalar wave parameters are within the corresponding white region of Fig.~\ref{aps}.

\begin{figure}[!htbp]
\begin{centering}
    \includegraphics[width=1.0\columnwidth]{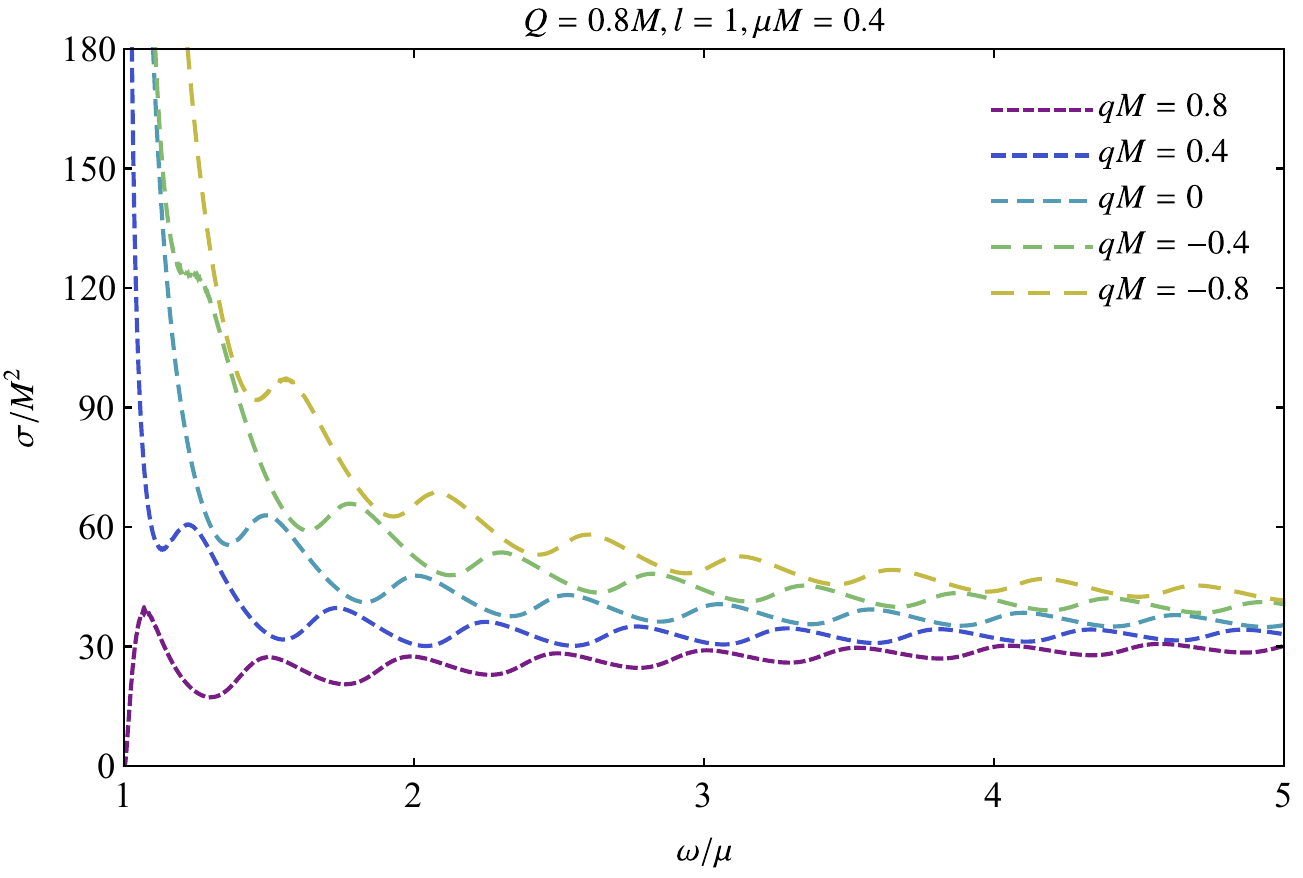}
    \caption{Total ACS of the scalar waves in the background of the charged bumblebee BH, as a function of $\omega / \mu$, for distinct values of $q M$ and considering $Q = 0.8M$, $l = 1$, and $\mu M = 0.4$.}
    \label{tacsdiffq}
\end{centering}
\end{figure}

In Fig.~\ref{tacsdiffq}, we display the total ACS of the charged bumblebee BH for distinct choices of $qM$. As we can see, the total ACS increases (diminishes) as we consider smaller values of $qQ$ for $q \phi_{+} > 0$ ($q \phi_{+} < 0$). This is related to the behavior of the potential barrier (see, e.g., the bottom panel of Fig.~\ref{potentialbarrier}) and can be understood by noting that the Lorentz force contributes to the repulsion (absorption) of scalar waves with the same (different) charge sign as the BH. Notice also that in Sec.~\ref{subsec:lfa}, we found that for $(1+l) ((2+l) q Q-2 (1+l) \mu  M) > 0$, the total ACS would be bounded and zero in the low-frequency regime [cf. Eq.~\eqref{lf_mM}]. For the parameters used in Fig.~\ref{tacsdiffl}, one can show that for $q \gtrsim 0.66$, the total ACS must satisfy Eq.~\eqref{lf_mM}, which is, in fact, what happens. Thus, the results shown in Fig.~\ref{tacsdiffq} once again highlight the agreement between our analytical and numerical results. We would also like to point out that this same analysis could be applied to other figures.

\begin{figure}[!htbp]
\begin{centering}
    \includegraphics[width=1.0\columnwidth]{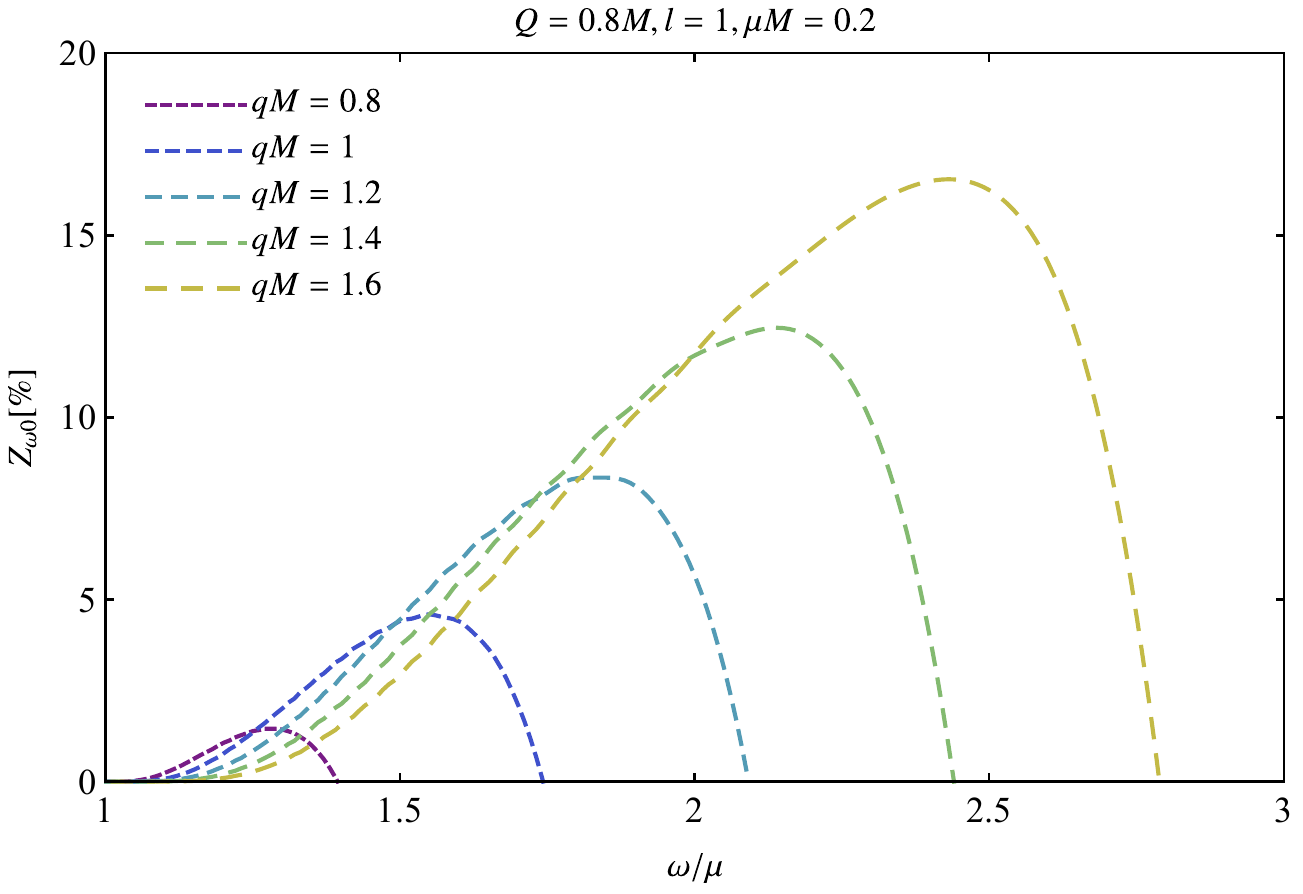}
    \includegraphics[width=1.0\columnwidth]{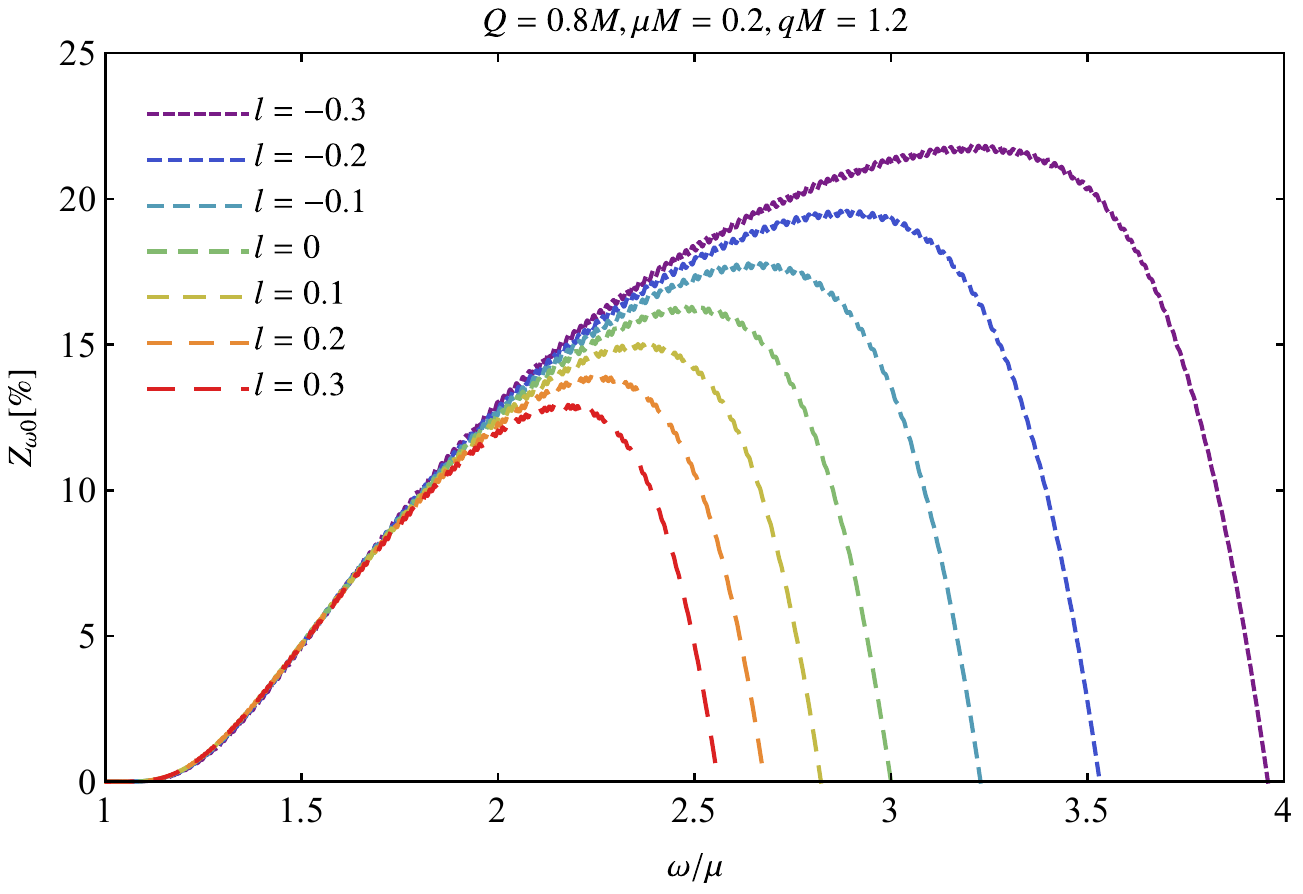}
    \caption{Superradiant amplification of the scalar waves in the background of the charged bumblebee BH, as a function of $\omega / \mu$, for two distinct scenarios: (i) different values of $q M$ with $Q = 0.8M$, $l = 1$, and $\mu M = 0.2$ (top panel); and (ii) distinct choices of $l$ with $Q = 0.8M$, $\mu M = 0.2$, and $q M = 1.2$ (bottom panel).}
    \label{ampfactorfig}
\end{centering}
\end{figure}

In Fig.~\ref{ampfactorfig}, we present the amplification factor of  charged massive scalar fields in the background of the charged bumblebee BH spacetime, considering distinct values of $qM$ and $l$. For simplicity, we exhibit the amplification factor in percentage, i.e., $Z_{\omega \ell}[\%] \equiv 100Z_{\omega \ell}$, and only consider the fundamental mode $\ell = 0$. We observe that the maximum superradiant amplification increases (decreases) as we increase the charge of the scalar field (the LV parameter), considering $q Q > 0$. This is consistent with the analyzes of the electrostatic potential (see, e.g., Fig.~\ref{epathorizon}) and the critical frequency [cf. Eq.~\eqref{criticalfreq}]. The superradiance threshold for charged BHs in bumblebee gravity increases as we consider higher values of $q M$ or $Q/M$, but decreases as we increase $l$. 

We emphasize that negative values of $l$ result in scalar waves with a higher amplification factor than in the RN case, as shown in Fig.~\ref{ampfactorfig}. Consequently, scalar waves in charged bumblebee BHs with $l < 0$ can extract more energy from the BH than in the RN case. This may play an interesting role in scenarios where scalar waves are confined. In such scenarios, since superradiant amplification in bumblebee gravity is greater than that in GR, we conjecture that superradiant instability is also greater.

\section{Conclusion}\label{sec:remarks}

We investigated the motion of charged particles and fields in the spacetime of a charged BH in bumblebee gravity, focusing on better understanding the role played by the LV parameter. We first reviewed the derivation of the BH solution and examined its physical and geometrical properties. For example, we obtained the event and Cauchy horizons, the extremal charge, the Kretschmann scalar, and the electrostatic potential. 

Regarding the motion of charged particles in charged bumblebee BHs, we derived the orbit equation through the Hamilton equations and analyzed the geometry of the orbits through the analysis of the turning points of the effective potential. Of particular interest are the circular curves associated with the motion of charged particles, found as critical points of the effective potential. The stability of these orbits is determined through the sign of the second derivative of the effective potential. This defines two branches of circular curves, namely the UCOs and the SCOs. These two branches intersect at the ISCO. Our results showed that the radius of the ISCO increases as the LV parameter increases, thus indicting that the distribution of accretion disks would change around different charged bumblebee BHs.

We also considered the most general bound orbits, particularly focusing on periodic orbits. Interestingly, bound orbits allowed in a charged bumblebee BH are not necessarily allowed in the
RN limit. We parametrized the orbits using the eccentricity and the latus rectum and provided a prescription of how to find  periodic orbits of charged particles in the charged bumblebee BHs. Following~\cite{lim2024energies,chan2025periodic} we derived the condition for the charged periodic orbits and studied them according to the Levin's taxonomy scheme~\cite{levin2008periodic}. In the $(L,E)$ space, the periodic orbits manifest as the so-called $\zeta$-branches with endpoints in the SCO branch and in the line $E=1$. We noticed that for smaller values of $l$ the $\zeta$-branches shift to the left, and consequently, for fixed BH charge,  the same family of periodic orbit requires larger energy and angular momentum in a BH with larger $l$.

Regarding the propagation of fields in the charged bumblebee BH, we investigated the absorption of charged massive scalar fields and, in particular, the occurrence of superradiance. Since in the charged bumblebee gravity, the Lorentz violation modifies the spherical sector of the metric, such that the equatorial plane acquires the geometry of a cone, we derived the total ACS by properly taking into account the non-flat asymptotics of the spacetime. We also derived low- and high-frequency approximations, which presents a very good agreement with our numerical results.

In particular, in the low-frequency regime, our approximation exhibits a transition between two branches, which is controlled by the velocity $v$. Specifically, on one hand, if $v\lesssim v_t$ one has to use the expression~\eqref{lf_1}; on the other hand, if $v\gtrsim v_t$ one has to use the expression~\eqref{lf_2}, where $v_t$ is the velocity of the transition given by Eq.~\eqref{vtrans}. In particular, these results reproduce the finite low-frequency cross section when Coulomb repulsion dominates over gravitational attraction [cf. Eq.~\eqref{criticalcase}]. In particular, we have shown that for $(1+l) ((2+l) q Q-2 (1+l) \mu  M) > 0$, the total ACS satisfies $\lim_{\omega \rightarrow \mu}\sigma = 0$. In the high-frequency regime, the ACS was obtained by analyzing the critical impact parameter of charged particles. Both approximations reduce to their RN counterparts when $l=0$.

It is well-established that in charged static BHs superradiance can occur if there is a feasible mechanism for the extraction of mass and charge from the BH. We investigated such a possibility through the analysis of the condition $\omega<\omega_{c}$, with $\omega_{c}=q\phi_{+}$. Thus, in our analysis the extraction mechanism is provided by the charge of the scalar field. Our results showed that for fixed positive values of $qM$ and $Q/M$, the radial electrostatic potential evaluated at the horizon decreases as $l$ increases, and consequently positive (negative) values of $l$ lead to narrower (wider) superradiant frequency intervals.

Our numerical results showed that the absorption parameter space displays unbounded absorption, bounded absorption, and bounded superradiance, as occurs for the RN case. Our results showed that although higher angular modes can carry a non-zero flux in the low-frequency regime for charged massive scalar waves, the amplitudes of these higher-order modes are heavily suppressed by the centrifugal barrier. Moreover, we  noticed that increasing $l$ reduces the total ACS even though it diminishes the height of the effective potential barrier. This reduction comes from the global factor $1/(1+l)$ associated with the non-flat asymptotic geometry. Therefore, although it is useful, analyzing the behavior of the potential is not necessarily sufficient to predict the behavior of the cross section, especially in spacetimes that are not asymptotically flat. Finally, the maximum superradiant amplification increased with the scalar field charge and decreased with $l$. 

\begin{acknowledgments}

The authors would like to acknowledge  Conselho Nacional de Desenvolvimento Cient\'ifico e Tecnol\'ogico (CNPq) from Brazil, for partial financial support. The authors would also like to thank Luís Carlos Bassalo Crispino for his helpful comments and discussions. A.A.A.F. is supported by  CNPq/PDJ 150223/2025-0. V. B. Bezerra is partially supported by CNPq through the Research Project No. 311847/2026-9. M. A. A. de Paula is supported by CNPq/PDJ 150589/2025-5. R. B. Magalhães is supported by CNPq/PDJ 151146/2025-0.

\end{acknowledgments}


\appendix

\section{Low-frequency limit of the cross section}\label{appx2}

\subsection{Revisiting the radial equations}\label{subsec:rre}

The metric function of the charged BH obtained in the bumblebee gravity can be written as
\begin{align}
\label{MF_bumb}f(r) = \left(1-\dfrac{r_{-}}{r}\right)\left(1-\dfrac{r_{+}}{r}\right) = \dfrac{h(r)}{1+l}.
\end{align}

However, as a solution to Eq.~\eqref{KG}, we write $\Phi$ as
\begin{equation}
\label{PHI}\Phi = \sum_{\ell =0}^{\infty} \psi_{\omega \ell}(r)P_{\ell}(\cos\theta)e^{-i\omega t},
\end{equation}
By inserting Eq.~\eqref{PHI} into Eq.~\eqref{KG}, we find that
\begin{equation}
\label{RE_RI}\dfrac{f(r)}{r^{2}}\left[r^{2}f(r)\psi_{\omega \ell }^{\prime}(r)\right]^{\prime}+U(r)\psi_{\omega \ell }(r) = 0,
\end{equation}
where the function $U(r)$ is given by
\begin{equation}
\label{effp2}U(r) = \left(1+l\right)\left\{ \left(\omega-q\phi(r)\right)^{2} - f(r)\left[\mu^{2}+ \dfrac{\ell(\ell+1)}{r^{2}}\right] \right\}.
\end{equation}

By defining a new radial function, namely
\begin{equation}
\label{new_rf}\psi_{\omega \ell }(r) \equiv \dfrac{\Psi_{\omega \ell }(r)}{r},
\end{equation}
and insert it into Eq.~\eqref{RE_RI}, we get
\begin{equation}
\label{RE_RII}\Psi_{\omega \ell}^{\prime\prime} (r) + \dfrac{f^{\prime}(r)}{f(r)}\Psi_{\omega \ell}^{\prime} (r) - \dfrac{\left(1+l\right)}{f(r)^{2}}V(r)\Psi_{\omega \ell}(r) = 0,
\end{equation}
where the function $V(r)$ is defined by Eq.~\eqref{EffP}. Considering the tortoise coordinate~\eqref{TC0}, we find that
\begin{equation}
\label{RE_RIII} \frac{\mathrm{d}^{2}}{\mathrm{d}r_{\star}^{2}}\Psi_{\omega \ell} - V(r)\Psi_{\omega \ell} = 0, 
\end{equation}
which is the same as Eq.~\eqref{RE}.

For the bumblebee BH case, the integration of $r_{\star}$ leads to
\begin{align}
\label{TC}r_{\star} = \ & \sqrt{1+l}\left[r + \dfrac{r_{+}^{2}}{r_{+}-r_{-}}\ln \left(r-r_{+}\right)- \right.\\
& \left. \dfrac{r_{-}^{2}}{r_{+}-r_{-}}\ln \left(r-r_{-}\right)\right] + r_{0},
\end{align}
where $r_{0}$ is an integration constant fixed, for simplicity, as
\begin{equation}
\label{radialconst}r_{0} = \sqrt{1+l}\left[\dfrac{r_{-}^{2}}{r_{+}-r_{-}}\ln \left(-r_{-}\right)-\dfrac{r_{+}^{2}}{r_{+}-r_{-}}\ln \left(-r_{+}\right)\right].
\end{equation}


\subsection{Low-frequency regime}

In this section, we focus on obtaining an analytic expression for the reflection coefficient in the low-frequency regime. For simplicity, we follow Refs.~\cite{Unruh:1976fm,Benone:2014qaa,Benone:2015bst,Richarte:2021fbi}. Therefore, we divide the spacetime into three distinct regions, namely:
\begin{itemize}
\item[(i)] \textit{Region I:} Very close to the BH event horizon;
\item[(ii)] \textit{Region II:} Intermediate region where the field is weak and we can take its low-frequency regime;
\item[(iii)] \textit{Region III:} Far away from the BH event horizon.
\end{itemize}


\subsubsection{Region I}

In \textit{Region I}, $r \rightarrow r_{+}$, and Eq.~\eqref{RE_RIII} can be written as
\begin{equation}
\label{RE_I} \dfrac{\mathrm{d}^{2}}{\mathrm{d}r_{\star}^{2}}\Psi_{\omega \ell}+\zeta^{2}\Psi_{\omega \ell}=0.
\end{equation}

According to the boundary conditions~\eqref{BC}, a possible solution of Eq.~\eqref{RE_I} is given by
\begin{equation}
\label{sol}\Psi_{\omega \ell } = T_{\omega \ell} e^{-i\zeta r_{\star}}.
\end{equation}
Considering Eq.~\eqref{new_rf}, we get
\begin{equation}
\label{truesol}\psi_{\omega \ell } = \dfrac{T_{\omega \ell} e^{-i\zeta r_{\star}}}{r}.
\end{equation}
In the limit $r \rightarrow r_{+}$, Eq.~\eqref{TC} reduces to
\begin{equation}
\label{TC_new}r_{\star} = \sqrt{1+l}\left[\dfrac{r_{+}^{2}}{r_{+}-r_{-}}\ln \left(r-r_{+}\right)\right] + r_{c},
\end{equation}
where $r_{c}$ is a constant defined as
\begin{equation}
\label{rc}r_{c} \equiv \sqrt{1+l}\left[r_{+} - \dfrac{r_{-}^{2}}{r_{+}-r_{-}}\ln \left(r_{+}-r_{-}\right)\right] + r_{0}.
\end{equation}

By inserting Eq.~\eqref{TC_new} into Eq.~\eqref{sol}, and recalling that $(1/r) \rightarrow (1/r_{+})$, in the limit $r\rightarrow r_{+}$, we get
\begin{equation}
\label{sol_rh}\psi^{(\rm{I})}_{\omega \ell}(r) = A_{\rm{TRA}}|r-r_{+}|^{-i\zeta \alpha},
\end{equation}
where $A_{\rm{TRA}}$ and $\alpha$ are defined as
\begin{align}
A_{\rm{TRA}} & \equiv \dfrac{T_{\omega \ell} e^{-i\zeta r_{c}}}{r_{+}}, \\
\alpha & \equiv \dfrac{\sqrt{1+l}\,r_{+}^{2}}{r_{+}-r_{-}},
\end{align}
respectively. We shall use the index $(\text{\textit{x}})$ to denote that the given solution is obtained in the \textit{Region x}.


\subsubsection{Region II}

In \textit{Region II}, we restrict our analysis to the low-frequency and low-mass limits ($\omega M \ll 1$ and $\mu M \ll 1$), as well as the weak electrostatic coupling limit ($qQ \ll 1$). In addition to that, for simplicity, we take into account only the contributions of the fundamental mode, i.e., we set $\ell = 0$. In this context, Eq.~\eqref{RE_RI} reduces to
\begin{equation}
\label{RE_RIInew}\psi_{\omega \ell}^{\prime\prime}(r) + \left(\dfrac{f^{\prime}(r)}{f(r)}+\dfrac{2}{r}\right)\psi_{\omega \ell}^{\prime}(r) = 0.
\end{equation}
For the bumblebee BH case, Eq.~\eqref{RE_RIInew} reads
\begin{equation}
\label{RE_RIInew_RN}\psi_{\omega \ell}^{\prime\prime}(r)+ \left(\dfrac{1}{r-r_{+}}+\dfrac{1}{r-r_{-}} \right)\psi_{\omega \ell}^{\prime}(r) = 0.
\end{equation}

The solution of Eq.~\eqref{RE_RIInew_RN} is given by
\begin{equation}
\psi_{\omega \ell}(r) = \dfrac{C_{1}}{r_{-}-r_{+}}\ln \left(\dfrac{r-r_{-}}{r-r_{+}}\right) + \tau,
\end{equation}
where $C_{1}$ and $\tau$ are integration constants. For simplicity, we rewrite $C_{1}$ as $C_{1} = \xi (r_{+}-r_{-})$. Then we get
\begin{equation}
\label{sol2}\psi^{(\rm{II})}_{\omega \ell}(r) = \xi \ln \left(\dfrac{r-r_{+}}{r-r_{-}}\right) + \tau,
\end{equation}
in which $\xi$ and $\tau$ are the constants to be determined.


\subsubsection{Region III}

In \textit{Region III}, $r \gg r_{+}$. In this context, considering Eq.~\eqref{RE_RII}, we can take into account only second-order contributions of the inverse radial coordinate and neglect terms of $\mathcal{O}(1/r^{2})$ proportional to $\mu^{2}$, $qQ$, $\omega^{2}$, and $r_{+}$. Consequently, we get
\begin{align}
\nonumber \Psi_{\omega \ell}^{\prime \prime}(r) + \bigg[&(1+l)\left(\omega^{2} - \mu^{2}\right) -\\
\nonumber & \dfrac{(l+2) q Q \omega + 2 M(l+1) \left(\mu ^2-2
   \omega ^2\right)}{r}-\\
\label{RE_III} & \dfrac{(l+1) \ell (\ell+1)}{r^2} \bigg]\Psi_{\omega \ell}(r) = 0.
\end{align}
By defining a new radial coordinate, namely
\begin{equation}
\label{newradial} y\equiv \sqrt{1+l} \,\kappa r,
\end{equation}
we can rewrite Eq.~\eqref{RE_III} as given by
\begin{equation}
\label{RE_IIInew} \dfrac{\mathrm{d}^{2}\Psi_{\omega \ell}}{\mathrm{d}y^{2}} + \left[1 -\dfrac{2\eta}{y}-\dfrac{\bar{\ell}\left(1+\bar{\ell}\right)}{y^{2}} \right]\Psi_{\omega \ell} = 0,
\end{equation}
which is a Coulomb wave equation. The complete solution is given by the well-known regular $F_{\bar{l}}(\eta, y)$ and irregular $G_{\bar{l}}(\eta, y)$  Coulomb functions~\cite{abramowitz1968handbook,barnett1996calculation,olver2010nist}. Therefore, considering Eq.~\eqref{new_rf}, we can write the solutions of Eq.~\eqref{RE_IIInew} as
\begin{equation}
\label{sol3}\psi_{\omega \ell }^{(\rm{III})}(r) = a\dfrac{F_{\bar{\ell}}(\eta, \sqrt{1+l}\kappa r)}{r} +b\dfrac{G_{\bar{\ell}}(\eta,\sqrt{1+l} \kappa r)}{r},
\end{equation}
where $a$ and $b$ are integration constants, with $\eta$ given by
\begin{equation}
\label{eta}\eta =\dfrac{ (2+l)qQ\omega + 2M(1+l)\left(\mu^{2}-2\omega^{2} \right)}{2\sqrt{1+l}\kappa},
\end{equation}
and $\bar{\ell}$ satisfies
\begin{equation}
\label{barl}(1+l)\ell(1+\ell) = \bar{\ell}(1+\bar{\ell}).
\end{equation}


\subsubsection{Matching the solutions}

First, we overlap \textit{Regions I} and \textit{II}. In the limit $r \rightarrow r_{+}$, Eqs.~\eqref{sol_rh} and~\eqref{sol2} can be written as
\begin{align}
\psi^{(\rm{I})}_{\omega \ell}(r) & \approx A_{\rm{TRA}}\left(1 -i\zeta \alpha \ln(r-r_{+}) \right), \\
\psi^{(\rm{II})}_{\omega \ell}(r) & \approx \xi \ln \left(r-r_{+}\right) - \xi \ln \left(r_{+}-r_{-}\right) + \tau.
\end{align}
By matching the coefficients, we get
\begin{align}
\label{xi}\xi & = - i\zeta \alpha A_{\rm{TRA}}, \\
\label{tauu}\tau & = (1-i \zeta \beta )A_{\rm{TRA}},
\end{align}
where
\begin{equation}
\label{beta}\beta \equiv \alpha \ln\left(r_{+}-r_{-}\right).
\end{equation}

Now, we overlap the \textit{Regions II} and \textit{III}. In the far field, Eq.~\eqref{sol3} leads to~\cite{barnett1996calculation,olver2010nist}
\begin{equation}
\label{sol3a}\psi_{\omega \ell }^{(\rm{III})}(r) = \dfrac{a\sin\vartheta_{\ell}}{r} + \dfrac{b\cos\vartheta_{\ell}}{r},
\end{equation}
where the function $\vartheta_{\ell}$ is defined as
\begin{align}
\nonumber \vartheta_{\ell} \equiv \ & \sqrt{1+l}\kappa r - \eta \ln(2\sqrt{1+l}\kappa r)+\\
\nonumber &\dfrac{\pi\left(1-\sqrt{4 (1+l) \ell (1+\ell)+1}\right)}{4}+\\
& \arg\Gamma\left(\dfrac{1}{2}+\dfrac{1}{2}\sqrt{4 (1+l) \ell (1+\ell)+1}+i\eta\right).
\end{align}
We can rewrite Eq.~\eqref{sol3a} as
\begin{equation}
\label{sol4}\psi_{\omega \ell }^{(\rm{III})}(r) = A_{\rm{INC}}\dfrac{e^{-i\vartheta_\ell}}{r}+A_{\rm{REF}}\dfrac{e^{i\vartheta_\ell}}{r},
\end{equation}
where
\begin{equation}
\label{coeff}A_{\rm{INC}} = \dfrac{-a+ib}{2i} \quad \text{and} \quad A_{\rm{REF}} = \dfrac{a+ib}{2i}.
\end{equation}

On the other hand, for $\sqrt{1+l}\kappa r \ll 1$, we get
\begin{equation}
\label{sol3b}\psi_{\omega \ell }^{(\rm{III})}(r) = \dfrac{aC_{\bar{\ell}}(\eta)\left(\sqrt{1+l}\kappa r\right)^{\bar{\ell}+1}}{r} + \dfrac{b\left(\sqrt{1+l}\kappa r \right)^{-\bar{\ell}}}{(2\bar{\ell}+1)C_{\bar{\ell}}(\eta)r},
\end{equation}
where $C_{\bar{\ell}}(\eta)$ is the so-called normalizing constant given by
\begin{equation}
\label{normalizationconst}C_{\bar{\ell}}(\eta) = \dfrac{2^{\bar{\ell}}}{(2\bar{\ell}+1)!}\left[\left(\dfrac{2\pi \eta}{e^{2\pi\eta}-1} \right)\prod_{k=1}^{\bar{\ell}}\left(\eta^{2}+k^{2} \right) \right]^{\frac{1}{2}}.
\end{equation}
Considering $\ell = 0$, Eq.~\eqref{sol3b} reduces to
\begin{equation}
\label{sol3c}\psi_{\omega \ell }^{(\rm{III})}(r) = a\rho \sqrt{1+l}\kappa + \dfrac{b}{\rho r},
\end{equation}
where we defined
\begin{equation}
\label{rho}\rho^{2} \equiv C_{0}(\eta)^{2} = \dfrac{2\pi \eta}{e^{2\pi \eta}-1}.
\end{equation}

Notice that in the limit $r \rightarrow \infty$, Eq.~\eqref{sol2} behaves as
\begin{equation}
\label{sol2a}\psi_{\omega \ell }^{(\rm{II})}(r) = -\xi \dfrac{\left(r_{+}-r_{-} \right)}{r} + \tau.
\end{equation}
By matching Eqs.~\eqref{sol3c} and~\eqref{sol2a} and using Eqs.~\eqref{xi} and~\eqref{tauu}, we find that
\begin{align}
\label{aa}a & = \dfrac{\left(1-i\zeta \beta \right)}{\rho \sqrt{1+l} \kappa}A_{\rm{TRA}}, \\
\label{bb}b & = i\rho \zeta \sqrt{1+l} r_{+}^{2}A_{\rm{TRA}}.
\end{align}
By inserting Eqs.~\eqref{aa} and~\eqref{bb} into Eqs.~\eqref{coeff}, we obtain
\begin{align}
\label{inccoeff}A_{\rm{INC}} = -\dfrac{1+(1+l)r_{+}^{2}\zeta \kappa \rho^{2}-i\beta \zeta}{2i \sqrt{1+l}\kappa\rho}A_{\rm{TRA}}, \\
\label{transcoeff}A_{\rm{REF}} = \dfrac{1-(1+l)r_{+}^{2}\zeta \kappa \rho^{2}-i\beta \zeta}{2i\sqrt{1+l}\kappa\rho}A_{\rm{TRA}}.
\end{align}

The reflection coefficient is defined as
\begin{equation}
\label{refcoef}|R_{\omega \ell}|^{2} \equiv \dfrac{|A_{\rm{REF}}|^{2}}{|A_{\rm{INC}}|^{2}},
\end{equation}
then we find that
\begin{equation}
\label{refcoef2}|R_{\omega \ell}|^{2} = \bigg|\dfrac{1-(1+l)r_{+}^{2}\zeta\kappa\rho^{2}-i\beta\zeta}{1+(1+l)r_{+}^{2}\zeta\kappa\rho^{2}-i\beta\zeta}\bigg|^{2}.
\end{equation}


\subsection{Low-frequency approximation}

By inserting Eq.~\eqref{refcoef2} into Eq.~\eqref{PACS}, and considering the fundamental mode, we obtain the analytical approximation for the low-frequency limit of the total ACS:
\begin{equation}
\label{lf}\sigma_{\rm{lf}} = \dfrac{\pi}{\kappa}\left(\dfrac{4 r_{+}^{2}\zeta \rho^{2}}{\left(1+(1+l)r_{+}^{2}\zeta \kappa \rho^{2}\right)^{2}+\beta^{2}\zeta^{2}} \right).
\end{equation}
To improve our understanding of the low-frequency approximation given by Eq.~\eqref{lf}, we can introduce the alternative dimensionless parameter $v$, namely
\begin{equation}
\label{speedofthefield}v = \dfrac{1}{\sqrt{1+l}}\sqrt{1-\dfrac{\mu^{2}}{\omega^{2}}},
\end{equation}
which corresponds to the ratio of the speed of propagation of the wave in the far field to the speed of light. Notice that since we are interested in unbounded modes, $v$ satisfies $0 < v \leq 1/\sqrt{1+l}$. Furthermore, the approximation given by Eq.~\eqref{lf} works well only for $\mu M \ll 1$ and $q Q \ll 1$, since we need these assumptions to obtain the corresponding radial equation in the intermediate region [cf. Eq.~\eqref{RE_RIInew}].


\section{Absorption parameter space}\label{appx}

In this appendix, we provide a simple but effective argument to construct and explain the absorption parameter space of charged BHs in bumblebee gravity. The boundary between bounded and unbounded absorption can be obtained from the asymptotic behavior of the potential function in the limit $\omega \rightarrow \mu$~\cite{dePaula:2024xnd,dePaula:2025kif}. Using Eq.~\eqref{EffP}, we find that
\begin{align}
\nonumber U(r) & \equiv -V(r)\big|_{\omega = \mu} \\
\label{EffP2}&= -\dfrac{\mu}{r} \left[2M\mu - \left(\dfrac{2+l}{1+l}\right)qQ \right] + \mathcal{O}\left[\dfrac{1}{r^{2}} \right].
\end{align}
At leading order, this expansion reduces to that of the RN BH when $l = 0$. We notice that for $2\mu M > (2+l)(1+l)^{-1}q Q$, the Newtonian attraction dominates over the Coulomb repulsion, and we have unbounded absorption (the total ACS diverges in the limit $\omega M \rightarrow \mu M$). Conversely, for $2\mu M < (2+l)(1+l)^{-1}q Q$, the Coulomb repulsion is dominant over the Newtonian attraction, and we have bounded absorption (the total ACS is finite in the limit $\omega M \rightarrow \mu M$). Moreover, the critical case occurs at
\begin{equation}
\label{criticalcase} \dfrac{\mu M}{qQ} = \dfrac{1}{2}\left(\dfrac{2+l}{1+l}\right).
\end{equation}

\begin{figure}[!htbp]
\begin{centering}
    \includegraphics[width=1.0\columnwidth]{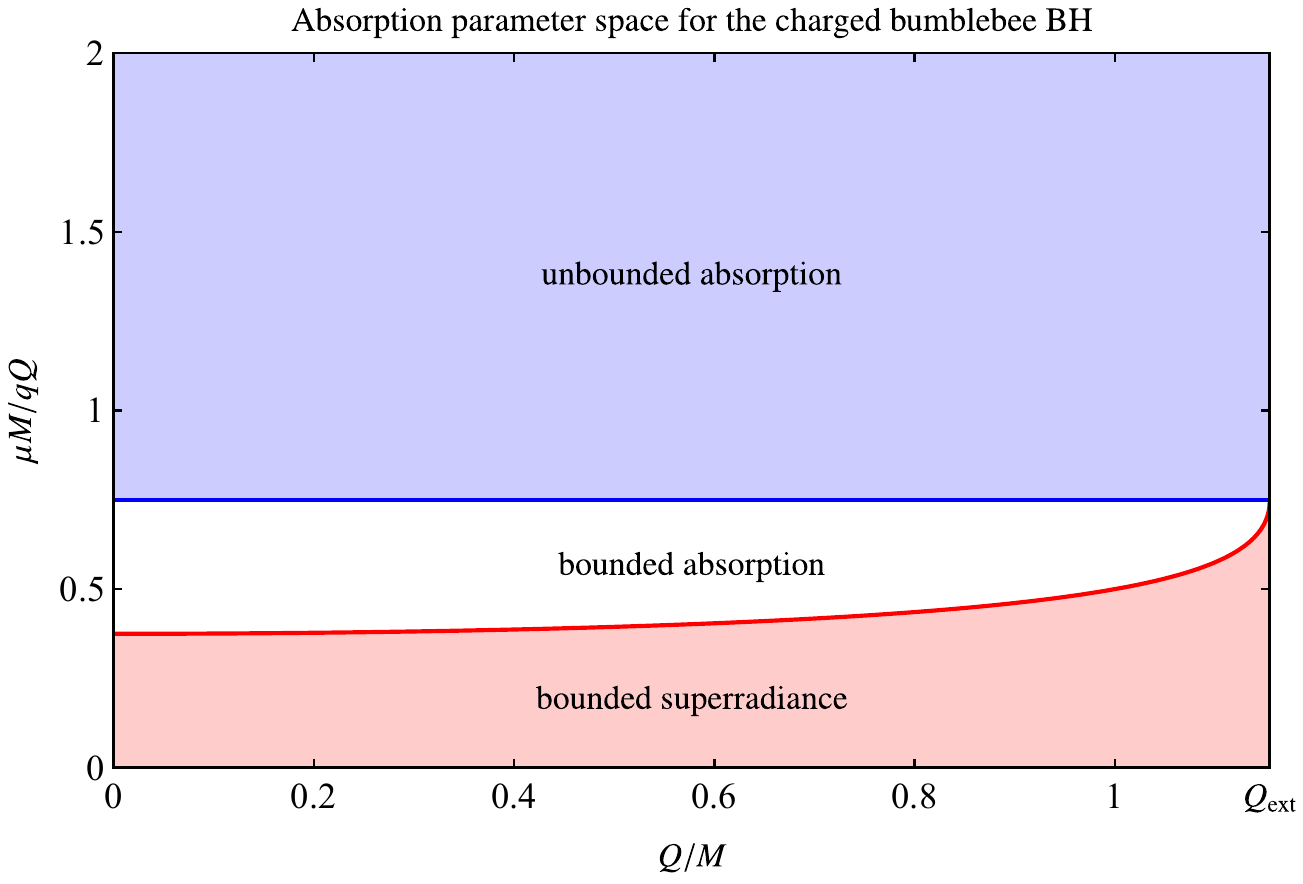}
    \caption{Absorption parameter space for the charged bumblebee BH, as a function of $Q/M$. The solid blue curve corresponds to the attractive/repulsive threshold [cf. Eq.~\eqref{criticalcase}] and the solid red curve to the superradiance threshold [cf. Eq.~\eqref{conditionsup}]. In this case, we set $l = 1$, for which the extremal charge is given by $Q_{\rm{ext}} = 1.1547M$.}
    \label{aps}
\end{centering}
\end{figure}

The threshold for superradiance is given by Eq.~\eqref{criticalfreq}, but in the limit $\omega \rightarrow \mu$, it can be written as
\begin{equation}
\label{conditionsup}\dfrac{\mu M}{qQ} = \dfrac{\phi_{+}M}{Q}.
\end{equation}
By combining Eqs.~\eqref{criticalcase} and~\eqref{conditionsup}, we can construct the absorption parameter space of charged bumblebee BHs, as shown in Fig.~\ref{aps}. Similar to the RN case, the charged bumblebee BH admits: (i) unbounded absorption; (ii) bounded absorption; and (iii) bounded superradiance. 

In Fig.~\ref{aps}, we fixed $l = 1$. To verify whether the scenario presented by this choice holds for any value of the LV parameter, we take the ratio between Eqs.~\eqref{criticalcase} and~\eqref{conditionsup}, given by
\begin{equation}
\label{ratio}\mathcal{Y} \equiv \dfrac{1}{2}\left(\dfrac{2+l}{1+l}\right)\dfrac{Q}{\phi_{+}M} = 1+\sqrt{1-\dfrac{(2+l) Q^2}{2 (1+l)M^{2}}}.
\end{equation}
Since the argument of the square root must always be greater than zero, we conclude that $\mathcal{Y} > 1$. Therefore, the generality of the parameter space shown in Fig.~\ref{aps} is still sound for any choice of $l$. We point out that this approach to constructing the absorption parameter space and ensuring its generality is complementary to the methods presented in Refs.~\cite{dePaula:2024xnd,dePaula:2025kif,Patrick:2021oqk}.

\bibliography{main}

\end{document}